\documentclass[epj,nopacs]{svjour}

\usepackage{amsmath}
\usepackage{amssymb}
\usepackage{mathtools}
\usepackage{latexsym}
\usepackage{amsfonts}
\usepackage{epsfig}
\usepackage{graphicx}
\usepackage{subfigure}
\usepackage{cite}

\usepackage{hyperref}

\usepackage[utf8]{inputenc}

\newcommand{\ket}[1]{\left|#1\right>}
\newcommand{\bra}[1]{\left<#1\right|}

\newcommand{\f}[1]{\mbox{\boldmath$#1$}}
\newcommand{\fk}[1]{\mbox{\boldmath$\scriptstyle#1$}}

\newcommand{\bea}{\begin{eqnarray}}
\newcommand{\ea}{\end{eqnarray}}
\newcommand{\eea}{\end{eqnarray}}
\newcommand{\ord}{\,{\cal O}}

\newcommand{\tr}{{\rm Tr}}

\newcommand{\cre}[1]{\hat{a}_{\fk{#1}}^\dagger}
\newcommand{\ann}[1]{\hat{a}_{\fk{#1}}^{\phantom{\dagger}}}

\newcommand{\tmn}{T_{\mu \nu}^{\phantom{\dagger}}}

\newcommand{\bcre}[1]{\hat{b}_{#1}^\dagger}
\newcommand{\bann}[1]{\hat{b}_{#1}^{\phantom{\dagger}}}
\newcommand{\ccre}[1]{\hat{c}_{#1}^\dagger}
\newcommand{\cann}[1]{\hat{c}_{#1}^{\phantom{\dagger}}}
\newcommand{\ka}{\fk{\kappa}}
\newcommand{\Ka}{\f{\kappa}}
\newcommand{\kin}{\fk{\kappa}_{\rm in}}
\newcommand{\Kin}{\f{\kappa}_{\rm in}}
\newcommand{\kout}{\fk{\kappa}_{\rm out}}
\newcommand{\Kout}{\f{\kappa}_{\rm out}}

\newcommand{\mot}{\rm Mo}
\newcommand{\nee}{\rm N\acute{e}}
\newcommand{\mon}{\rm Mo/N\acute{e}}
\newcommand{\exc}{\rm ex}
\newcommand{\sfl}{\rm sf}
\newcommand{\mtl}{\rm me}

\newcommand{\se}{s_1\hspace{-0.5pt}}
\newcommand{\sz}{s_2}
\newcommand{\sd}{s_3}
\newcommand{\sv}{s_4}

\begin{document}

\title{Dicke superradiance as nondestructive probe for the state of atoms in 
optical lattices}

\author{Nicolai ten Brinke\thanks{\email{nicolai.ten-brinke@uni-due.de}} \and Ralf Sch\"utzhold\thanks{\email{ralf.schuetzhold@uni-due.de}}}

\institute{Fakult\"at f\"ur Physik, Universit\"at Duisburg-Essen, 
Lotharstra\ss e 1, D-47057 Duisburg, Germany}

\date{\today}

\abstract{
We present a proposal for a probing scheme utilizing Dicke superradiance 
to obtain information about ultracold atoms in optical lattices.
A probe photon is absorbed collectively by an ensemble of lattice atoms
generating a Dicke state. 
The lattice dynamics (e.g., tunneling) affects the coherence properties of 
that Dicke state and thus alters the superradiant emission characteristics 
-- which in turn provides insight into the lattice (dynamics).
Comparing the Bose-Hubbard and the Fermi-Hubbard model, we find similar 
superradiance in the strongly interacting Mott insulator regime, but crucial 
differences in the weakly interacting (superfluid or metallic) phase. 
Furthermore, we study the possibility to detect whether a quantum phase 
transition between the two regimes can be considered adiabatic or a quantum 
quench.
}

\authorrunning{ten Brinke \and Sch\"utzhold}

\maketitle

\section{Introduction}
\label{sec:introduction}
%
Optical lattices are artificial crystals of light, created by interfering 
optical laser beams~\cite{Bloch:2004eg}.
Storing ultracold bosonic or fermionic atoms in optical lattices allows the 
creation of model systems in which many parameters can easily be controlled, 
including the periodic structure and the interaction between the 
atoms~\cite{Raizen:1997yq,Bloch:2004eg,Bloch:2005zl,Bloch:2008aa}.
Remarkably, the virtual absence of decoherence and temperature effects
in an isolated optical lattice enables a direct view on quantum 
many-body physics, thus rendering ultracold atoms 
in optical lattices a versatile tool~\cite{Jaksch:2005kq} 
to be used in a wide range of experiments.
One of the phenomena which can be studied particularly well is
quantum phase transitions~\cite{Sachdev:2001kx}.
In the case of bosonic atoms in an optical lattice, for example,
a phase transition between the superfluid phase, described by 
a macroscopic wave function spreading throughout the entire lattice, 
and the Mott insulator phase, corresponding to a fixed number of atoms
per lattice site, is predicted by the Bose-Hubbard
model~\cite{Fisher:1989fv,Jaksch:1998uq,Jaksch:2005kq,Krutitsky:2015aa}.
This phase transition, induced by varying the depth of the 
optical lattice potential, was observed experimentally%
~\cite{Greiner:2002fk,Stoferle:2004qv,Spielman:2007kx} via 
time-of-flight measurements, in which the atoms are
abruptly released from the lattice potential and their positions
are then detected via absorption imaging.
A matter-wave interference pattern featuring sharp
peaks is obtained for the phase coherent superfluid state%
~\cite{Greiner:2002fk}, while their absence
indicates a Mott insulating state%
\footnote{
This is a simplified view, however,
as it was recently pointed out~\cite{Kato:2008vn}, 
that sharp peaks do not always pinpoint superfluidity.}.
Alternatively, the superfluid to Mott insulator transition was also
detected via microwave spectroscopy~\cite{Campbell:2006aa}
or~Bragg scattering \cite{Miyake:2011aa}.
In the case of fermions, on the other side,
the Fermi-Hubbard model~\cite{Hubbard:1963aa,Jaksch:2005kq}
describes a metal-insulator transition, which has 
been extensively studied in the field of 
condensed-matter systems~\cite{Mott:1968aa,Imada:1998aa}.
While experimentally more challenging than in the 
case of bosons, the Fermi-Hubbard model was recently
implemented in a three-dimensional optical lattice%
~\cite{Kohl:2005aa,Esslinger:2010aa}.
In latest experiments, it was possible to detect metallic
and insulating phases~\cite{Schneider:2008aa} and to
observe the formation of a fermionic Mott insulator%
~\cite{Jordens:2008aa}.
Notably, time-of-flight measurements directly image
the Fermi surface of the atoms in the optical lattice,
as initial momentum maps into final position%
~\cite{Kohl:2005aa,Esslinger:2010aa}.
As opposed to measuring the momentum distribution
of the lattice atoms, direct in situ imaging
makes it possible to reconstruct the atom number
distribution in the optical lattice with single-atom and
single-site resolution, as realized in, e.g.,%
~\cite{Schneider:2008aa,Gemelke:2009aa,Itah:2010aa,
Sherson:2010uq,Endres:2013fk,Greif:2016aa}.
Although there is no doubt that the aforementioned 
methods offer extremely valuable insights,
they suffer from an obvious drawback: 
the quantum state of the atoms in the optical lattice
is destroyed or collapsed (loss of phase coherence)
by the measurement.
Therefore, other less destructive probing schemes
were proposed, 
such as off-resonant collective light scattering from 
the atoms trapped in the optical lattice%
~\cite{Mekhov:2007aa,Mekhov:2007ac,Mekhov:2007ab,Chen:2007aa},
described via a two-band Bose-Hubbard model coupled to
cavity light fields%
~\cite{Bhaseen:2009fv,Zoubi:2009ty,Silver:2010rz,Rajaram:2013zl}.
Following this approach, measuring the light scattered into the 
optical cavity should allow to distinguish between 
different atomic quantum phases.
For fermionic atoms in optical lattices, on the other hand, 
an all-optical pump-and-probe scheme to obtain information
about dynamical (two-time) correlations
was proposed recently~\cite{Dao:2010aa}.
Experimentally, vac\-u\-um-stim\-u\-lat\-ed scattering of light 
was employed to nondestructively 
measure the dynamic structure factor of a quantum gas~\cite{Landig:2015aa}.
Aside from light scattering, other approaches to nondestructively
probe quantum phase transitions in an optical lattice are matter-wave 
scattering with (slow) atoms, as proposed in~\cite{Sanders:2010aa,
Mayer:2014aa}, and the interference of the trapped atoms with a 
reference Bose-Einstein condensate~\cite{Niu:2006aa}.

In this paper, we extend the probing scheme first presented
in \cite{Brinke:2015aa} to a general formalism, 
placing special emphasis on the difference between bosonic 
and fermionic systems.
We describe in detail an alternative, nondestructive detection
method which is based on Dicke superradiance, i.e., 
the collective and coherent absorption and (time delayed) 
emission of photons from an ensemble of ultracold atoms%
~\cite{Dicke:1954kx,Rehler:1971fk,Lipkin:2002fk,
Akkermans:2008aa,Wiegner:2011aa}.
To infer the quantum state of the optical lattice
from the emission characteristics,
we analyze in which way the lattice dynamics (e.g., tunneling),
developing after the absorption of a single photon, 
alters its subsequent superradiant emission%
~\cite{Scully:2006fk,Scully:2007fk,Oliveira:2014aa,Bromley:2016aa}.
As opposed to the instantaneous off-resonant 
Bragg-type scattering into a cavity%
~\cite{Mekhov:2007aa,Mekhov:2007ac,Mekhov:2007ab,Chen:2007aa}, 
we study resonant Dicke superradiance in free space with a time delay 
in between absorption and emission\footnote{
We recently became aware of the pump-and-probe detection
scheme presented in reference~\cite{Dao:2010aa}, which shares some
key ingredients with our approach, e.g.\ the storage of a coherent
light pulse, its decoherence due to lattice dynamics and its later retrieval
which gives insight into the atomic two-time correlations.
However, instead of focusing on fermions and the BCS superfluid state, 
we here develop a general method based on Dicke superradiance
which can be applied to both bosonic and fermionic lattices in a variety 
of scenarios. 
}.
Because of the time delay, our approach corresponds to pump-probe
spectroscopy in analogy to solid state physics. Therefore, it provides
important complementary information, e.g., for non-equilibrium
phenomena.
For instance, our method is sensitive to the correlator of creation 
and annihilation operators including their phase coherence at 
different times~(see Eq.~(\ref{eq:OpD}) below), as in e.g.%
~\cite{Niu:2006aa,Dao:2010aa}, instead of the 
correlator containing on-site number operators at the 
same time only, as in references%
~\cite{Mekhov:2007aa,Mekhov:2007ac,Mekhov:2007ab}.

\begin{figure}[t]
\begin{center}
\includegraphics[width=0.9\columnwidth]{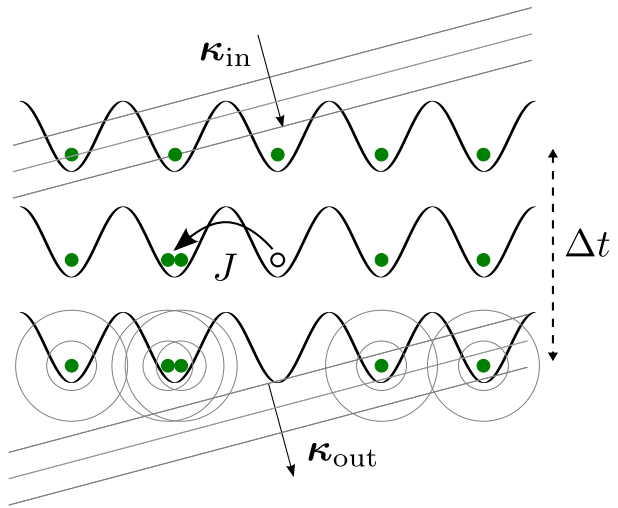}
\caption{
Envisaged probing sequence: 
first, a probe photon $\Kin$ is absorbed collectively
by the ground-state atoms at their respective lattice sites.
Second, the (bosonic or fermionic) atoms tunnel and interact 
according to, e.g., the Bose- or Fermi-Hubbard model 
during a waiting period $\Delta t$, thereby possibly compromising
the spatial phase coherence of the Dicke state.
In the third step, the probe photon is (collectively) emitted again with the
wave vector $\Kout$ -- where the emission characteristics were probably 
altered due to the lattice dynamics.
}
\label{fig:probe_dicke}
\end{center}
\end{figure}

Before we start with a detailed explanation of our model
in the upcoming sections, let us briefly describe
the general idea behind our probing scheme, as shown
in Figure~\ref{fig:probe_dicke}.
The probing sequence consists of three steps.
At first, the probe photon is sent almost (but not quite) orthogonally onto 
the two dimensional optical lattice, where it is absorbed by one of the 
lattice atoms.
Assuming that the recoil of the probe photon is small enough such that it is 
transferred to the optical lattice, we cannot know which atom absorbed the 
photon.
Thus, a coherent superposition state -- a ``timed'' Dicke 
state~\cite{Scully:2006fk,Scully:2007fk,Oliveira:2014aa} --
is created. 
In a second step, the usual lattice dynamics (tunneling, interaction), 
as e.g.\ described by the Bose- or Fermi-Hubbard model, 
evolves during a waiting period $\Delta t$.
Thirdly, the atoms collectively re-emit the previously absorbed
probe photon. 
However, the tunneling of the lattice atoms
altered the phase coherence of the Dicke state,
thus resulting in modified emission characteristics,
as compared to the case of immovable atoms.
In this paper, we try to answer the question what
can be learned about the quantum state 
(or about phase transitions) of the (bosonic or fermionic)
optical lattice by observing these emission characteristics.

Next (Sect.~\ref{sec:model}), we introduce the basic model
for our calculations, in particular, the Hamiltonians 
of the lattice dynamics and interaction with the probe photon,
followed by a brief introduction to superradiance.
In Section~\ref{sec:EmissionProb} we derive a general
expression for the emission probability, 
which is then further investigated for different
parameter regimes of the optical lattice in 
Section~\ref{sec:SeparableState} (separable state)
and Section~\ref{sec:WeakInteractions} (weak interactions).
The general findings are applied to concrete
examples in terms of the Bose- and Fermi-Hubbard model, 
when we discuss which lattice states can be 
distinguished via our probing scheme
in Section~\ref{sec:ProbingLatticeStates}.
Going further, we inspect what signs of quantum phase
transitions can be detected by our probe in Section%
~\ref{sec:Transitions}.
We discuss several options for experimental realization
in Section~\ref{sec:Experimental} and another particular 
option in Section~\ref{sec:Classical}, which employs 
classical laser fields instead of the absorption and 
spontaneous emission of a single photon.
Finally, we conclude in Section~\ref{sec:Conclusions}.
%
\section{Model}
\label{sec:model}
%
\subsection{Hamiltonian}
The basic idea of this paper is to investigate optical lattices
via the interaction of the lattice atoms with infrared photons.
Hence, we introduce the following general lattice Hamiltonian.
It consists of a tunneling term with the tunneling rate $J$ and an 
on-site repulsion term with the interaction strength $U$,
\begin{align}
\label{eq:GenHam}
\hat{H}_{\rm lat} &=  -\frac{J}{Z}\sum_{\mu\nu, s,\lambda}\tmn \hat{a}_{\mu s}^{\lambda\,\dagger} \hat{a}_{\nu s}^{\lambda}\nonumber\\
&\hspace{0.4cm}+ \frac{U}{2}\sum_\mu \Bigg( \sum_{s,\lambda} \hat{n}_{\mu s}^{\lambda} \Bigg)\Bigg( \sum_{s,\lambda} \hat{n}_{\mu s}^{\lambda}-1 \Bigg)
\,,
\end{align}
and describes the dynamics of cold atoms in an optical lattice, 
assuming lowest band occupation only and neglecting
long-range interactions%
\footnote{
Longer range (e.g. dipole-dipole) interactions are usually much 
weaker than the short range (on-site) forces, and are therefore 
neglected.
Note that though ground and excited state atoms in general
behave differently with regard to dipole-dipole interactions,
the resulting phase factors would still be negligible in comparison
with the phase factors stemming from the tunneling rate $J$,
which are responsible for the effects discussed in this paper.
}.
Furthermore, the adjacency matrix $T_{\mu\nu}$ and
the coordination number $Z$ are determined by
the lattice structure. In our case, we assume a
two-dimensional, quadratic lattice with 
$Z=\sum_\nu \tmn=4$ and $N$ lattice sites and atoms.
Above all, $\hat{a}_{\mu s}^{\lambda\,\dagger}$ and 
$\hat{a}_{\nu s}^{\lambda}$ denote (bosonic or fermionic) 
creation and annihilation operators of atoms at lattice sites 
$\mu$ and $\nu$, respectively. 
Their number operator is abbreviated as $\hat{n}_{\mu s}^{\lambda}$.
In order to model the absorption and emission of photons
by the atomic ensemble, we need to distinguish between
(at least) two internal levels $\lambda \in \{ {\rm gr, \rm ex} \}$ 
of the atoms.
In addition, the index $s$ refers to a potential extra spin quantum
number -- if applicable, such as for fermionic atoms.
Note that we assume identical lattice dynamics~(\ref{eq:GenHam})
for both the ground and excited state atoms, for simplicity.
However, our qualitative results do not depend on this presupposition%
\footnote{
Although we assume that ground and excited state atoms underlie
the same tunneling rate $J$ and on-site repulsion $U$, the qualitative
results of this paper hold true in the general case.
In the separable state regime ($U\gg J$), for example, 
the important point is that the atoms are (almost) fixed to their 
lattice sites -- which is also true for differing interaction strengths,
as long as they are all large compared to the tunneling rate $J$.
In the weak interactions regime ($J\gg U$), on the other hand,
different $J$ would appear in the formulae. But in the end, 
there are still the same distinguishable cases -- 
either the usual superradiance or a decay of the superradiance peak.
}.
The complete Hamiltonian of our system, including the general lattice
dynamics as in equation~(\ref{eq:GenHam}) as well as 
the free-space electromagnetic field and the
interaction between atoms and photons, is given by ($\hbar=1$)
\begin{align}
\label{eq:Hsystem}
\hat{H}= \hat{H}_{\rm lat} + \omega \sum_{\mu,s} \hat{n}_{\mu s}^{{\rm ex}} +\int d^3k\, \omega_{\fk{k}}^{\phantom{\dagger}} \hat{a}_{\fk{k}}^\dagger\hat{a}_{\fk{k}}^{\phantom{\dagger}}+ \hat{V}
\,,
\end{align}
where the middle two terms account for the energy stored in the excited
atoms (the energy gap between the two atomic levels is denoted by $\omega$),
and the energy stored in the electromagnetic field, respectively.
Here, $\ann{k}$ and $\cre{k}$ are the usual annihilation and creation
operators for photons with wave number $\f{k}$.
The internal states of the atoms are coupled to the electromagnetic
field via the perturbation Hamiltonian in rotating-wave and dipole
approximation,
\begin{align}
\label{eq:hInt}
\hat{V} = 
\int d^3k\, g_{\fk{k}}^{\phantom{\dagger}}(t) \ann{k} 
\hat{\Sigma}^+\left(\f{k}\right)
+ {\rm H.c.}
\,,
\end{align}
where the (time-dependent) coupling constant $g_{\fk{k}}(t)$ is 
assumed to be small. Note that it is not 
space-dependent as in e.g.~\cite{Dao:2010aa}, that is we assume 
equal coupling of all atoms.
Most important, the exciton creation 
operator is given by:
\begin{align}
\label{eq:ExCre}
\hat{\Sigma}^+(\f{k}) = \sum_{\mu,s} \hat{a}_{\mu s}^{{\rm ex}\,\dagger} \hat{a}_{\mu s}^{{\rm gr}}\exp\left\{i\f{k}\cdot\f{r}_\mu\right\}
\,,
\end{align}
where $\f{r}_\mu$ is the position of the atom at the lattice site 
$\mu$.
Each summand describes the excitation of an (bosonic or fermionic)
atom from the ground state to the excited state with the 
corresponding spatial phase factor. The sum over all lattice sites
creates a (entangled, i.e., W-type) superposition state -- 
in other words, we do not know which
atom has been excited.
Note that the spin quantum number $s$ of the lattice atoms 
should not affect the hopping, interaction or excitation process, 
as considered via the summation over $s$
in equations~(\ref{eq:GenHam}), (\ref{eq:Hsystem}) and~(\ref{eq:ExCre}).
\subsection{Dicke superradiance}
From the exciton creation operator 
$\hat{\Sigma}^+(\f{k})$ and its Hermitian conjugate counterpart,
the exciton annihilation operator 
$\hat{\Sigma}^-(\f{k}) = [\hat{\Sigma}^+(\f{k})]^\dagger$,
quasispin-$N$-operators $\hat{\Sigma}^x(\f{k})$, $\hat{\Sigma}^y(\f{k})$
and $\hat{\Sigma}^z$ can be constructed via
$\hat{\Sigma}^\pm(\f{k})=\hat{\Sigma}^x(\f{k})\pm i \hat{\Sigma}^y(\f{k})$
and 
\begin{align}
\hat{\Sigma}^z 
= 
\frac{1}{2}\left[ \hat{\Sigma}^+(\f{k}), \hat{\Sigma}^-(\f{k}) \right]
= 
\frac{1}{2}\sum_{\mu,s}\left( \hat{n}_{\mu s}^{{\rm ex}}-\hat{n}_{\mu s}^{{\rm gr}} \right)
\,.
\end{align}
The quasispin-$N$-operators generate an $SU(2)$ algebra~\cite{Lipkin:2002fk,Akkermans:2008aa},
which leads to the astonishing effect that the transition probabilities 
roughly grow with the number $N$ of atoms \emph{times} the number $n$ of 
excitations
\begin{align}
\label{eq:DickeMatrix}
\hat{\Sigma}^+(\f{k})\ket{n}&=\sqrt{(N-n)(n+1)}\ket{n+1}\,,\nonumber\\
\hat{\Sigma}^-(\f{k})\ket{n}&=\sqrt{(N-n+1)n}\ket{n-1}
\,,
\end{align}
where $\ket{n}\propto [\hat{\Sigma}^+(\f{k})]^n\ket{0}$ denotes a coherent
superposition state with $n$ excitons, often referred to as Dicke state%
~\cite{Dicke:1954kx}.
Apart from the increased absorption and emission probability,
another important consequence of the collective effects
when an \emph{ensemble} of atoms absorbs a photon 
is directed spontaneous emission~\cite{Scully:2006fk,Scully:2007fk,
Oliveira:2014aa,Bromley:2016aa}.
In short, after the absorption of a photon with wave vector
$\f{\kappa}$, the atom-ensemble re-emits the photon
predominantly in the same (forward) direction $\f{\kappa}$.
Intuitively speaking, the wave vector $\f{\kappa}$
of the absorbed photon is remembered by the
atom-ensemble via the spatial phases $\exp\left(i\f{\kappa}\cdot\f{r}_\mu\right)$
in the Dicke state, stemming from exciton creation~(\ref{eq:ExCre}).
In the following, we will restrict ourselves to \emph{single-photon} 
superradiance, where according to equation~(\ref{eq:DickeMatrix}), 
the absorption ($n=0$) and directed spontaneous emission ($n=1$) 
probabilities are enhanced by a factor $N$ each.

However, certain requirements have to be met in order to
reach superradiance.
For instance, superradiant is decay is dominant over incoherent 
emission processes when the lattice spacing $\ell$ of the optical lattice
is small compared to the probe-photon wavelength $\lambda$%
~\cite{Brinke:2013fk}.
Recent calculations and experiments~\cite{Bettles:2015aa,
Bromley:2016aa} suggest that it is sufficient when the lattice
spacing is only slightly smaller (but still of the same order)
than the driving wavelength.
As another requirement, the atomic recoil due to the absorption 
or emission of the probe photon should be negligible.
For both reasons, it is favorable to consider
infrared photons. We will revisit this issue when discussing
options for experimental realization in Section~\ref{sec:Experimental}.
%
\section{Emission Probability}
\label{sec:EmissionProb}
%
In the formalism outlined above, let us now calculate the probability density
for the event that a photon with wave vector $\Kin$ is absorbed and 
subsequently re-emitted with wave vector $\Kout$ after a waiting time 
$\Delta t$.
Employing first-order perturbation theory for the absorption
as well as for the emission process, the probability density reads
\begin{align}
\label{eq:ProbDens1}
P
=\left\|\bra{0}\hat{a}_{\kout}\int_0^{\tau_E}\hspace{-0.2cm}dt_2\,\hat{V}(t_2)\int_0^{\tau_A}\hspace{-0.2cm}dt_1\,\hat{V}(t_1)
\ket{\Psi}\hat{a}_{\kin}^\dagger\ket{0}\right\|^2
\,.
\end{align}
Here, the $\ket{0}$-ket refers to the vacuum state of the photon field,
while $\hat{V}(t)$ denotes the perturbation Hamiltonian~(\ref{eq:hInt})
in the interaction picture.
The initial (bosonic or fermionic) state of the optical lattice $\ket{\Psi}$ 
can be arbitrary, except that we assume that there are no 
$\hat{a}_{\mu s}^{{\rm ex}\,\dagger}$-excitations in the beginning, 
i.e., $\hat{a}_{\mu s}^{{\rm ex}}\ket{\Psi}=0$ for all $\mu, s$.
Resolving the photon part yields an expression containing only the 
exciton creation operator~(\ref{eq:ExCre}) and its adjoint
$( \hat{\Sigma}^+)^\dagger =: \hat{\Sigma}^-$ as operatorial part,
\begin{align}
\label{eq:ProbDens2}
P=\mathcal{I}\left[ e^{i(\omega_{\rm out}t_2-\omega_{\rm in}t_1)}\hat{\Sigma}^-(\Kout,t_2) \hat{\Sigma}^+(\Kin,t_1)\ket{\Psi}\right]
\,,
\end{align}
where, in order to simplify notation, we introduced a functional 
$\mathcal{I}[f]$ as abbreviation 
\begin{align}
\label{eq:ProbDensFunc}
\mathcal{I}\left[ f \right] = \bigg\|\int_0^{\tau_E}\hspace{-0.2cm}dt_2\int_0^{\tau_A}\hspace{-0.2cm}dt_1\, g_{\kout}^*(t_2) g_{\kin}(t_1) f(t_1,t_2)\bigg\|^2
\,.
\end{align}
For further analysis of the operatorial part,
we expand the probability density as a scalar product,
\begin{multline}
\label{eq:ProbDensScalar}
P=\int dt_1\,dt_2\,dt_3\,dt_4\, g_{\kout}(t_4) g_{\kout}^*(t_2) g_{\kin}^*(t_3) g_{\kin}(t_1)\\
\times\,e^{i(\omega_{\rm in}t_3-\omega_{\rm out}t_4)}
e^{-i(\omega_{\rm in}t_1-\omega_{\rm out}t_2)} 
\mathcal{D}\left(t_1, t_2, t_3, t_4\right)
\,,
\end{multline}
and break it down to the individual lattice sites~\cite{Brinke:2015aa},
\begin{align}
\label{eq:OpD}
\mathcal{D}\left(t_1, t_2, t_3, t_4\right)
&=\sum_{\mu\nu\rho\eta,\se\sz\sd\sv}\hspace{-0.35cm} \exp\left\{i\left(\Kout\cdot\f{r}_\rho-\Kin\cdot\f{r}_\eta\right)\right\}\nonumber\\
&\hspace{1.25cm}\times\exp\left\{-i\left(\Kout\cdot\f{r}_\mu-\Kin\cdot\f{r}_\nu\right)\right\}\nonumber\\
&\hspace{-1.2cm}\times\bra{\Psi} \hat{a}_{\eta \sv}^{{\rm gr}\,\dagger}(t_3) \hat{a}_{\eta \sv}^{{\rm ex}}(t_3) \hat{a}_{\rho \sd}^{{\rm ex}\,\dagger}(t_4) \hat{a}_{\rho \sd}^{{\rm gr}}(t_4)\nonumber\\
&\hspace{-0.65cm}\times \hat{a}_{\mu \se}^{{\rm gr}\,\dagger}(t_2)\hat{a}_{\mu \se}^{{\rm ex}}(t_2)\hat{a}_{\nu \sz}^{{\rm ex}\,\dagger}(t_1)\hat{a}_{\nu \sz}^{{\rm gr}}(t_1)\ket{\Psi}
\,.
\end{align}
Note that while we presented the derivation for a pure state $\ket{\Psi}$
as the initial state of the optical lattice, we could as well start with an arbitrary
mixed state $\hat{\rho}_{\rm in}$, for example a thermal state, instead. 
In this case, the expectation value $\bra{\Psi}\hat{A}\ket{\Psi}$
in the lower part of equation~(\ref{eq:OpD}) is replaced by the corresponding
expression for a mixed state, i.e., $\tr\{ \hat{\rho}_{\rm in}\hat{A}\}$.
In any case, the lattice dynamics is incorporated 
in the operatorial function $\mathcal{D}$ in 
equation~(\ref{eq:OpD}) via the time-dependency 
of the annihilation and creation operators.
In addition, the collective behavior of the absorption
and emission is encoded in the summation of the
spatial phases.
In summary, all interesting phenomena are guided
by the four-times eight-point function in the lower 
part of equation~(\ref{eq:OpD}). 
Unfortunately, it cannot be solved explicitly for the 
general lattice Hamiltonian~(\ref{eq:GenHam})
without further assumptions.
In the upcoming sections, we will thus study the limiting cases of separable
states (i.e., small $J$) and weak interactions (i.e., $J\gg U$).
%
\section{Separable state}
\label{sec:SeparableState}
%
Let us first have a look at the case of negligible correlations between the 
lattice sites. 
One example is the Mott insulator state (for $U\gg J$).
Note that one should carefully distinguish correlations between lattice sites
on the one hand from correlations between particles on the other hand. 
For example, the Mott insulator state is often called strongly correlated 
because it displays strong correlations between particles.
However, the correlations between lattice sites become negligible for 
$U\gg J$. 

We assume that the initial state $\ket{\Psi}$ can be written as a product of 
(bosonic or fermionic) normalized single-site states, i.e., we employ
the Gutzwiller ansatz~\cite{Gutzwiller:1963aa,Queisser:2014aa},
\begin{align}
\ket{\Psi}=\bigotimes_\mu \ket{\Psi_\mu} = \bigotimes_\mu \left(\ket{\psi_\mu}^{{\rm gr}}\otimes\ket{0_\mu}^{{\rm ex}}\right)
\,.
\end{align}
Obviously, there are zero correlations and
no $\hat{a}_{\mu s}^{{\rm ex}\,\dagger}$-ex\-ci\-ta\-tions at the beginning.
Further, we set $t=t_{1/3}$ as the time at which the 
photon is absorbed and $t'=t_{2/4}$ as the time at which the photon
is emitted again. 
This is justified, as we require that the waiting time $\Delta t = t'-t$
between the absorption and the emission of the photon is much larger
than the time taken by the absorption or emission process itself,
because the lattice dynamics is typically much slower. 
The correlator from~(\ref{eq:OpD}) then reads
\begin{multline}
\label{eq:Corr1}
\hspace{-0.2cm}\mathcal{C}^{\mu\nu\rho\eta}_{\se\sz\sd\sv}\left(t,t'\right)
=
\big(\otimes_{\xi} \bra{\Psi_\xi}\big)\hat{a}_{\eta \sv}^{{\rm gr}\,\dagger}(t) \hat{a}_{\eta \sv}^{{\rm ex}}(t) \hat{a}_{\rho \sd}^{{\rm ex}\,\dagger}(t') \hat{a}_{\rho \sd}^{{\rm gr}}(t')\\
\times \hat{a}_{\mu \se}^{{\rm gr}\,\dagger}(t')\hat{a}_{\mu \se}^{{\rm ex}}(t')\hat{a}_{\nu \sz}^{{\rm ex}\,\dagger}(t)\hat{a}_{\nu \sz}^{{\rm gr}}(t)
\big(\otimes_{\chi} \ket{\Psi_\chi}\big)
\,.
\end{multline}
Furthermore, we require that the correlations which arise during the waiting 
time $\Delta t$ remain negligible.
Under these assumptions, equation~(\ref{eq:Corr1}) can be 
evaluated for different cases regarding the indices.
It turns out that to leading order in $N$, there is only
one case where the correlator does not vanish (for details,
see Appendix~\ref{ap:DCorrelators}),
\begin{multline}
\hspace{-0.2cm}\mathcal{C}^{\mu\mu\eta\eta}_{\se\sz\sd\sv}\left(t,t'\right)
=
\bra{\Psi_\eta}\hat{a}_{\eta \sv}^{{\rm gr}\,\dagger}(t) \hat{a}_{\eta \sv}^{{\rm ex}}(t) \hat{a}_{\eta \sd}^{{\rm ex}\,\dagger}(t') \hat{a}_{\eta \sd}^{{\rm gr}}(t')\ket{\Psi_\eta}\\
\times\bra{\Psi_\mu}\hat{a}_{\mu \se}^{{\rm gr}\,\dagger}(t')\hat{a}_{\mu \se}^{{\rm ex}}(t')\hat{a}_{\mu \sz}^{{\rm ex}\,\dagger}(t)\hat{a}_{\mu \sz}^{{\rm gr}}(t)\ket{\Psi_\mu}
\,.
\end{multline}
Inserting back into the operatorial part~(\ref{eq:OpD}),
and contracting the sums according to $\mu=\nu$ and $\eta=\rho$,
the result can be written as an absolute square~\cite{Brinke:2015aa}
\begin{multline}
\label{eq:DLowCorr}
\mathcal{D}\left(t,t'\right)
= \Bigg|\sum_{\mu, \se\sz} \exp\left\{-i\left(\Kout-\Kin\right)\cdot\f{r}_\mu\right\}\\
\times\bra{\Psi_\mu}\hat{a}_{\mu\se}^{{\rm gr}\,\dagger}(t')\hat{a}_{\mu\se}^{{\rm ex}}(t')\hat{a}_{\mu\sz}^{{\rm ex}\,\dagger}(t)\hat{a}_{\mu\sz}^{{\rm gr}}(t)\ket{\Psi_\mu}\Bigg|^2
\,.
\end{multline}
The result for the operatorial function for separable
states (\ref{eq:DLowCorr}), is quite intuitive.
It corresponds to the probability amplitude that the
excited atom which is created at lattice site $\mu$
(with spin $\sz$) is still at the same position $\mu$
(with spin $\se$) after the waiting time $\Delta t = t'-t$.
Going to the extreme case of $J=0$, i.e., the case
of immovable atoms, the time-dependency of the creation
and annihilation operators can be calculated explicitly
and cancels out (up to a global temporal phase).
Without $\hat{a}_{\mu s}^{{\rm ex}\,\dagger}$-excitations 
in $\ket{\Psi_\mu}$, the operatorial part~(\ref{eq:DLowCorr}) reduces to
\begin{align}
\mathcal{D}\left(t,t'\right) 
=
\left|\sum_{\mu} \exp\left\{-i\left(\Kout-\Kin\right)\cdot\f{r}_\mu\right\} n_\mu
 \right|^2
\,,
\end{align}
where $n_\mu$ is just the total number of (ground-state) atoms 
at lattice site $\mu$, i.e., $n_\mu = 
\sum_{s}\bra{\Psi_\mu}\hat{n}_{\mu s}^{{\rm gr}}\ket{\Psi_\mu}$.
In other words, in the case $J=0$, the sum can be understood
as a discrete Fourier transform of the $n_\mu$-distribution
of the atoms in the optical lattice.
An obvious example is a state with one atom per lattice
site, $n_\mu = 1$, which yields a sharp peak
$\mathcal{D}\left(t,t'\right) = N^2 \delta_{\kin\kout}$
from the Fourier transform%
\footnote{Where we always use the atomic summation
for a large number density, i.e., 
$\sum_{\mu} \exp\left\{-i\left(\Kout-\Kin\right)\cdot\f{r}_\mu\right\}
=
N \delta_{\kin\kout}$.}.
As in this case the atoms are fixed to their lattice
sites, it is evident that we reproduce the well-known
directed spontaneous superradiant emission%
~\cite{Scully:2006fk,Scully:2007fk,Oliveira:2014aa},
\begin{align}
\label{eq:Pmott}
P = N^2 \delta_{\kin\kout} P_{\rm single}
\,,
\end{align}
where $P_{\rm single}$ is the emission probability
density for a single atom 
(for details, see Appendix~\ref{ap:ProbSingle}).
Note that only one factor of $N$ originates from
the (single-photon) superradiant emission, 
while another factor $N$
stems from the simple fact that $N$ atoms absorb the
incident probe photon more likely than one atom.
However, keep in mind that~(\ref{eq:DLowCorr})
is only valid in the case of negligible correlations.
Turning this argument around, a deviation from this behavior 
is then an indication for correlations. 
In Section~\ref{sec:ProbingLatticeStates} we will see clear deviations
from equation~(\ref{eq:DLowCorr}).
%
\section{Weak interactions}
\label{sec:WeakInteractions}
%
Now we study the opposite limit, $J\gg U$, which typically features 
strong correlations between the lattice sites 
(think, e.g., of the superfluid state of the Bose-Hubbard model).
Approximating $U=0$, the general lattice Hamiltonian%
~(\ref{eq:GenHam}) becomes diagonal in the $\f{k}$-basis 
(for further details, see Appendix~\ref{ap:kBasis}).
Hence, the exciton creation operator~(\ref{eq:ExCre}) 
picks up a phase 
\begin{align}
\label{eq:InterPhase}
\phi_{\fk{p}}^{\fk{k}}(t) = -J/Z(T_{\fk{p}}-T_{\fk{p}-\fk{k}})t
\,,
\end{align}
in the interaction picture, i.e.\footnote{
Please note that we tacitly required here
that the wave vector $\f{k}$ is a lattice vector, 
and will continue to do so for any wave vector
from now on.
Otherwise, expressions would get quite lengthy
and provide less insight, although our main results
hold true.
Here, for example, the $\mu$-sum in~(\ref{eq:ExCre}) 
would not yield an exact Kronecker delta after the transformation
to $\f{k}$-space.
However, in the experiment, the wave vector $\Kin$ 
of the probe photon can simply be tuned such that 
it matches a lattice vector (Sect.~\ref{sec:Experimental}).},
\begin{align}
\label{eq:ExCreIntPic}
\hat{\Sigma}^+\left(\f{k}, t\right) 
= 
e^{i\omega t}\sum_{\fk{p}, s}\hat{a}_{\fk{p}, s}^{{\rm ex}\,\dagger}\hat{a}_{\fk{p}-\fk{k}, s}^{{\rm gr}}\exp\left\{i\phi_{\fk{p}}^{\fk{k}}(t)\right\}
\,,
\end{align}
where $T_{\fk{p}}$ denotes the Fourier transform of the
adjacency matrix $T_{\mu \nu}$.
After inserting the exciton creation~(\ref{eq:ExCreIntPic}) 
and annihilation operator (its adjoint) into the probability 
density~(\ref{eq:ProbDens2}),
we reduce the expression via 
$\hat{a}_{\fk{k},\se}^{{\rm ex}}\hat{a}_{\fk{p},\sz}^{{\rm ex}\,\dagger}\ket{\Psi} 
= 
\delta_{\fk{k}\fk{p}}\delta_{\se\sz}\ket{\Psi}$,
i.e., we employ (anti-)commutation relations and 
include the fact that $\hat{a}_{\mu s}^{{\rm ex}}\ket{\Psi}=0$.
Moreover, we assume that the incoming photon
is in resonance with the atomic transition, i.e.,
$\omega_{\rm in}=\omega$. 
When we expand the probability density 
analogous to the lattice-site basis as in
equations~(\ref{eq:ProbDensScalar}) and~(\ref{eq:OpD}),
we eventually find
\begin{multline}
\label{eq:ProbDensHigh}
P=
\int dt_1\,dt_2\,dt_3\,dt_4\,
g_{\kout}(t_4) g_{\kout}^*(t_2) 
g_{\kin}^*(t_3) g_{\kin}(t_1) 
\\
\times\,e^{-i(\omega_{\rm out}-\omega)t_4}
e^{i(\omega_{\rm out}-\omega)t_2}
\mathcal{E}\left(t_1, t_2, t_3, t_4\right)
\,,
\end{multline}
with the operatorial part $\mathcal{E}$,
which once again contains the interesting
phenomena,
\begin{align}
\label{eq:OpE}
\mathcal{E}\left(t_1, t_2, t_3, t_4\right)
&=
\sum_{\fk{k}\fk{q},\se\sz}\hspace{-0.1cm}
\exp\left\{i\left[\phi_{\fk{q}}^{\kout}(t_4)-\phi_{\fk{q}}^{\kin}(t_3)\right]\right\}\nonumber\\
&\hspace{0.9cm}\times
\exp\left\{-i\left[\phi_{\fk{k}}^{\kout}(t_2)-\phi_{\fk{k}}^{\kin}(t_1)\right]\right\}\nonumber\\
&\hspace{-2.0cm}\times
\bra{\Psi}
\hat{a}_{\fk{q}-\kin,\sz}^{{\rm gr}\,\dagger}\hat{a}_{\fk{q}-\kout,\sz}^{{\rm gr}} \hat{a}_{\fk{k}-\kout,\se}^{{\rm gr}\,\dagger}\hat{a}_{\fk{k}-\kin,\se}^{{\rm gr}}
\ket{\Psi}
\,.
\end{align}
Again, the result is also valid for a mixed state $\hat{\rho}_{\rm in}$
instead of a pure state $\ket{\Psi}$ as the initial state of the optical lattice,
when we substitute $\tr\{ \hat{\rho}_{\rm in}\hat{A}\}$ for
the expectation value $\bra{\Psi}\hat{A}\ket{\Psi}$.
Note that so far, we did not specify whether we deal with bosonic or 
fermionic lattice atoms. 
However, as we will dive into two concrete examples in the 
upcoming Section~\ref{sec:ProbingLatticeStates} (namely the Bose-
and Fermi-Hubbard model), we will differentiate between these 
two setups from now on.
Regarding the four-point correlator in equation~(\ref{eq:OpE}), we consider
two cases in which it can be explicitly calculated.

First and foremost, if the initial pure state $\ket{\Psi}$ is diagonal in the
$\f{k}$-basis, we can define the eigenvalues of the number operator
as $\hat{n}_{\fk{k}, s}^{{\rm gr}}\ket{\Psi} = n_s(\f{k}) \ket{\Psi}$, and
express the four-point correlator in equation~(\ref{eq:OpE}) in terms of these
eigenvalues $n_s(\f{k})$.
In the case of bosons without spin quantum number 
($\se,\sz$ omitted), the correlator then yields
\begin{align}
\label{eq:BosonicCorr}
E_n^{\rm B} &= \bra{\Psi}
\hat{a}_{\fk{q}-\kin}^{{\rm gr}\,\dagger}\hat{a}_{\fk{q}-\kout}^{{\rm gr}} \hat{a}_{\fk{k}-\kout}^{{\rm gr}\,\dagger}\hat{a}_{\fk{k}-\kin}^{{\rm gr}}
\ket{\Psi}\nonumber\\
&=
n(\f{k}-\Kin) n(\f{q}-\Kout)\left(\delta_{\kin\kout}+\delta_{\fk{k}\fk{q}}\right)\nonumber\\
&\hspace{0.4cm}+n(\f{k}-\Kin)\delta_{\fk{k}\fk{q}}\nonumber\\
&\hspace{0.4cm}-n(\f{k}-\Kin) \left[ n(\f{q}-\Kout) + 1 \right] \delta_{\fk{k}\fk{q}}\delta_{\kin\kout}
\,,
\end{align}
while in the case of fermions with internal spin 
$\se,\sz \in \left\{ \uparrow,\downarrow \right\}$,
the correlator gives
\begin{align}
\label{eq:FermionicCorr}
E_n^{\rm F}
&= 
\bra{\Psi}\hat{a}_{\fk{q}-\kin,\sz}^{{\rm gr}\,\dagger}\hat{a}_{\fk{q}-\kout,\sz}^{{\rm gr}} \hat{a}_{\fk{k}-\kout,\se}^{{\rm gr}\,\dagger}\hat{a}_{\fk{k}-\kin,\se}^{{\rm gr}}
\ket{\Psi}\nonumber\\
&=
n_{\se}(\f{k}-\Kin) n_{\sz}(\f{q}-\Kout) \left( \delta_{\kin \kout} - \delta_{\fk{k} \fk{q}}\delta_{\se\sz} \right)\nonumber\\
&\hspace{0.4cm}+
n_{\se}(\f{k}-\Kin) \delta_{\fk{k} \fk{q}} \delta_{\se\sz}
\,.
\end{align}
For details on the calculations, see Appendix~\ref{ap:FourPointCorr}. 

As an alternative, we can utilize Wick's theorem 
\cite{Wick:1950aa,Evans:1996aa} for Gaussian states $\hat{\rho}_{\rm g}$,
such as thermal states. Wick's theorem states that the
four-point correlator in equation (\ref{eq:OpE}) can be expanded into a sum 
of products of two-point functions, when the Hamiltonian is quadratic
in the annihilation and creation operators (which is the case for weak 
interactions, i.e., $U=0$). 
The resulting two-point functions can then be expressed as the
expectation values of the number operators in the Gaussian state,
i.e., $n_s(\f{k})=\tr\{ \hat{\rho}_{\rm g}\hat{n}_{\fk{k}, s}^{{\rm gr}}\}$.
After applying Wick's theorem, the result for the bosonic case
is given by the first two lines of equation~(\ref{eq:BosonicCorr}) -- the
last line does not appear in this case, as it corresponds to the
situation where all four creation and annihilation operators act
on the same $\f{k}$-mode.
Note that this discrepancy is insignificant, as the term in the
last line is usually negligible compared to the first term.
In the fermionic case, Wick's theorem yields the same 
expression~(\ref{eq:FermionicCorr}) as for a 
diagonal pure state $\ket{\Psi}$.

We now have everything at hand to calculate the emission
characteristics~(\ref{eq:ProbDens1}) via~(\ref{eq:ProbDensHigh})
in the limit of weak interactions ($J\gg U$) for a diagonal pure state
$\ket{\Psi}$ or a Gaussian state $\hat{\rho}_{\rm g}$.
The remaining task is to insert the number
distribution (in reciprocal space)
$n_{s}(\f{k})$ of the state,
e.g., the number of particles per mode $\f{k}$, 
in equation~(\ref{eq:BosonicCorr}) or~(\ref{eq:FermionicCorr}),
respectively.
Thus, in the next section, we will examine concrete
lattice states in the context of the Bose- and Fermi-Hubbard model.
%
\section{Probing lattice states}
\label{sec:ProbingLatticeStates}
%
In this section, the general results above are applied to the
two most commonly discussed optical lattice models --
the Bose-Hubbard model and the Fermi-Hubbard
model.
To this end, a brief introduction on the respective
model (and its ground states in different regimes) 
is given at the beginning of each subsection.
\subsection{Bose-Hubbard model}
\begin{figure}[t]
\begin{center}
\includegraphics[width=1.0\columnwidth]{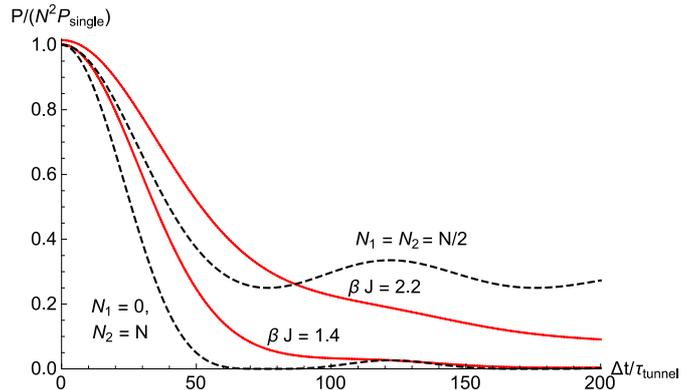}
\caption{ 
Emission probability $P$ from equation~(\ref{eq:ProbDensHigh})
(normalized by $N^2 P_{\rm single}$, where one factor $N$ originates
from single-photon superradiance and another factor $N$ stems
from the absorption process) for emission
in direction $\Kout=\Kin=2\pi / (L \ell)\left\{1,1\right\}$ 
over the waiting time $\Delta t$ in units of the tunneling time 
$\tau_{\rm tunnel}=\hbar/J$ for $N=L^2=100^2$ 
lattice sites and bosons.
The dashed black lines represent numerical results
for the partial condensation state~(\ref{eq:probability}),
while the solid red lines represent numerical results for the
Bose-Einstein distribution~(\ref{eq:BEdist}).
Note that for high temperatures, $\beta J \ll 1$, the result for
the Bose-Einstein distribution matches the result of the
totally distributed case, as depicted by the lower
dashed line.
}
\label{fig:decay_bos}
\end{center}
\end{figure}

On the one hand, 
bosonic ultracold atoms in optical lattices can approximately be described
by the Bose-Hubbard Hamiltonian%
~\cite{Fisher:1989fv,Jaksch:1998uq,Jaksch:2005kq,Krutitsky:2015aa}. 
The two-species Bose-Hubbard Hamiltonian is a special case of the 
general lattice Hamiltonian (\ref{eq:GenHam}), describing
the tunneling $J$ and on-site repulsion $U$ of bosons 
with no spin quantum number
(i.e., index $s$ omitted),
\begin{align}
\label{eq:BHHamExt}
\hat{H}_{\rm BH}^{\rm 2sp}&= -\frac{J}{Z}\sum_{\mu\nu, \lambda}\tmn\hat{b}_{\mu}^{\lambda\,\dagger} \hat{b}_{\nu}^{\lambda}\nonumber\\
&\hspace{0.4cm}+\frac{U}{2} \sum_\mu \left( \sum_{\lambda} \hat{n}_\mu^{\lambda}\right) \left( \sum_{\lambda} \hat{n}_\mu^{\lambda} - 1 \right)
\,,
\end{align}
where $\hat{b}_{\mu}^{\lambda\,\dagger}$ and $\hat{b}_{\nu}^{\lambda}$
are now \emph{bosonic} creation and annihilation operators, of course.
As we require that there are no excited bosons initially
(and there is no way to excite them except the probe photon),
the spectrum of possible initial lattice states is 
determined by the usual, single-species Bose-Hubbard
Hamiltonian (i.e., index $\lambda$ omitted),
\begin{align}
\label{eq:BHHam}
\hat{H}_{\rm BH}=  -\frac{J}{Z}\sum_{\mu\nu}\tmn\bcre{\mu}\bann{\nu} + \frac{U}{2} \sum_\mu \hat{n}_\mu ( \hat{n}_\mu - 1 )
\,.
\end{align}
In the following, we will assume unit filling, 
$\langle\hat{n}_\mu\rangle=1$,
i.e., one boson per lattice site, on average.
In this case, the Bose-Hubbard model features
a quantum phase transition~\cite{Sachdev:2001kx}
between the Mott insulator state, where the interaction
term dominates, $U\gg J$, and the superfluid phase,
where the tunneling term dominates, $J\gg U$.
\setcounter{paragraph}{0}
\paragraph{Mott insulator state}
In the first extremal limit, $U \gg J$, the atoms are pinned
to their lattice sites and the ground state is given by the
fully localized Mott insulator state,
\begin{align}
\label{eq:MottBos}
\ket{\Psi}^{J=0}_{\mot}=\prod\limits_\mu\bcre{\mu}\ket{0}=
\bigotimes\limits_\mu\ket{1}_\mu
\,.
\end{align}
As discussed in Section~\ref{sec:SeparableState},
we obtain the usual single-photon superradiance in this case,
i.e., the emission probability of the probe photon
is enhanced by a factor $N^2$ as compared to the
single atom case~(\ref{eq:Pmott}).
\paragraph{Superfluid ground state}
In the other extremal limit, $J\gg U$, the atoms 
can move freely across the entire lattice, which leads
to the completely delocalized superfluid ground state,
\begin{align}
\label{eq:SuperfluidBos}
\ket{\Psi}_{\sfl}^{U=0}
=
\frac{1}{\sqrt{N!}}
\big(\hat b_{\fk{k}=0}^\dagger\big)^N\ket{0}
\propto
\left(\sum\limits_\mu\bcre{\mu}\right)^N\ket{0}
\,.
\end{align}
In this case, the number distribution in 
Section~\ref{sec:WeakInteractions} collapses to
$n^{\sfl}(\f{k})=N \delta_{\fk{k}0}$,
which yields for the bosonic four-point correlator%
~(\ref{eq:BosonicCorr})
\begin{align}
\label{eq:CorrSfGround}
E_{n^{\sfl}}^{\rm B}
=
N \left(N-1\right)\delta_{\fk{k}\fk{q}\kin\kout}
+
N\delta_{\fk{k}\fk{q}\kin}
\,.
\end{align}
When we insert this result into the
operatorial part~(\ref{eq:OpE}), 
all sums are fixed and we obtain a term
proportional to $N^2$ to leading order.
Its global phase $\exp\{i \phi_{\kin}^{\kin}(t_4-t_3-t_2+t_1)\}$
actually depends on the waiting time $\Delta t$ 
(see Sect.~\ref{sec:SeparableState}),
i.e., the time in between absorption and emission.
As the waiting time $\Delta t$ is assumed to
be much larger than the timescales of the
absorption and emission process itself,
the phase can be regarded as constant over
the integration periods. Thus we arrive at
\begin{align}
\label{eq:Psfgs}
P = N^2 \delta_{\kin\kout} P_{\rm single} + \ord(N)
\,,
\end{align}
where, again, $P_{\rm single}$ is the emission probability
density for a single atom.
Comparing~(\ref{eq:Pmott}) and~(\ref{eq:Psfgs}), 
the two extremal bosonic ground states%
~(\ref{eq:MottBos}) and~(\ref{eq:SuperfluidBos})
show the same (single-photon) superradiant emission 
characteristics (to leading order in $N$),
and are thus not distinguishable using our probe.
\paragraph{Partial condensation state}
Therefore we regarded a more general state in reference%
~\cite{Brinke:2015aa} which contains a certain number $N_1$
of bosons in the ground state at $\f{k}=0$ while the other $N_2$ 
bosons should be equally distributed between all $\f{k}$-modes.
Such a state could, for example, represent a simple toy model
for a thermal state with partial condensation.
Starting from the corresponding number distribution
$n^{\rm dt}(\f{k}) = N_1\delta_{\fk{k}\fk{0}} + N_2/N$,
we obtained the emission probability density (for a detailed 
derivation, see Appendix~\ref{ap:PartialCond})
\begin{align}
\label{eq:probability}
P=\left|N_1 e^{i\varphi(\Delta t)}+N_2{\mathcal J}(\Delta t)\right|^2
\delta_{\kin\kout}P_{\rm single}
\,.
\end{align}
Of course, for $N_1=N$ and $N_2=0$, the result
of the superfluid ground state~(\ref{eq:Psfgs})
is reproduced.
Much more interesting is the opposite scenario of
a state where all $\f{k}$-modes are equally populated
by $N_2>0$ bosons.
In this case, the emission probability decays
over the waiting time $\Delta t$ due to the
phase-sum factor ${\mathcal J}(\Delta t)$ in equation%
~(\ref{eq:reduction}) in Appendix~\ref{ap:PartialCond},
except for an orthogonal probe photon with
$\kappa_x = \kappa_y = 0$.
The decay of the (single-photon) superradiance peak for
$N_2=N$ and $N_2=N/2$ is 
shown in Figure~\ref{fig:decay_bos} 
(dashed black lines). 
The dependence on the lattice vector becomes
more clear when we approximate the 
phase-sum $\mathcal{J}(\Delta t)$
via Bessel functions $J_0$, which is justified
if the wave number is small compared to the
lattice spacing $|\f{\kappa}|\ell\ll1$, and the
number of lattices sites in one dimension is
large, $L\gg1$
(see Appendix~\ref{ap:approx_phasesum}),
\begin{align}
\label{eq:bessels}
\mathcal{J}(\Delta t)
\approx
J_0 \left( 2\frac{J\Delta t}{Z}\,\kappa_x\ell \right) 
J_0 \left( 2\frac{J\Delta t}{Z}\,\kappa_y\ell \right) 
\,.
\end{align}
We find that the magnitude of the reduction is guided
by $J \Delta t$, determining the amount of
tunneling, times the wave-vector
(components) $\kappa_{x/y}$.
Qualitatively, this can be understood
in the following picture.
The decay of the superradiance peak is 
caused by the lattice dynamics during the waiting
time $\Delta t$, in which the (excited) atoms tunnel 
according to the tunneling rate $J$ and thus
corrupt the spatial phase coherence which is 
crucial for superradiance.
A comparable effect -- the reduction of superradiance
due to motional effects -- was recently observed 
experimentally for a dense coherent medium~\cite{Bromley:2016aa}.
The damage inflicted on the phase
coherence, however, also depends on how much the spatial
phases of neighboring atoms differ --
which is guided by the wave vector $\f{\kappa}$.
In general, larger wave-vector components $\kappa_{x/y}$, 
i.e., a more skewed entrance angle of the probe photon, 
leads to a more rapid decay of the superradiance peak. 
In the extreme case of an orthogonal probe photon,
$\kappa_x = \kappa_y = 0$, the spatial phases 
$\exp\left(i\f{\kappa}\cdot\f{r}_\mu\right)$
do not differ at all and thus, there is no reduction.
Note that the result~(\ref{eq:bessels}) stands in opposition
to the result we obtained for separable states~(\ref{eq:DLowCorr}),
e.g., by its explicit dependence on the wave vector $\f{\kappa}$,
thus demonstrating the importance of the correlations which
arise due to the tunneling of the atoms.
\paragraph{Thermal state}
Instead of the above toy model state
with partial condensation, let us now
insert the more realistic Bose-Einstein distribution,
\begin{align}
\label{eq:BEdist}
n^{\rm BE}(\f{k}) = \frac{1}{e^{\beta\left(E_{\fk{k}}-\mu\right)}-1}
\,,
\end{align}
where, according to the Bose-Hubbard model~(\ref{eq:HJextk}),
the energy of a particle in the mode $\f{k}$ is 
given by $E_{\fk{k}}=-J/Z\, T_{\fk{k}}$ (for $U=0$).
In this case, the expressions~(\ref{eq:BosonicCorr})
and~(\ref{eq:OpE}) get quite lengthy, although 
they are still given analytically, of course.
In particular, the phase-sums over $\f{k}$ and $\f{q}$
cannot be simplified, e.g., via Bessel-functions,
as in the partial condensation state~(\ref{eq:probability}).
Thus, we show numerical results for the
emission probability~(\ref{eq:ProbDensHigh}),
where the Bose-Einstein distribution was
inserted in equation~(\ref{eq:BosonicCorr}), in 
Figure~\ref{fig:decay_bos} (solid red lines).

Let us briefly sum up the results for a bosonic lattice.
For the Mott state and the superfluid ground state,
we obtained the usual single-photon superradiance 
-- the result is independent of the waiting time $\Delta t$, 
implying the lattice dynamics in between absorption
and emission does not change the 
emission characteristics.
For (partially) excited states such as the toy-model
state~(\ref{eq:probability}) or the more realistic
Bose-Einstein distribution~(\ref{eq:BEdist}), 
the superradiance peak decays.
In conclusion, bosonic (partially) excited states,
such as, e.g., thermal states,
can be distinguished from the superfluid 
ground state via Dicke superradiance.

\subsection{Fermi-Hubbard model}
\label{sec:FermiHubbard}
\begin{figure}[t]
\begin{center}
\includegraphics[width=1.0\columnwidth]{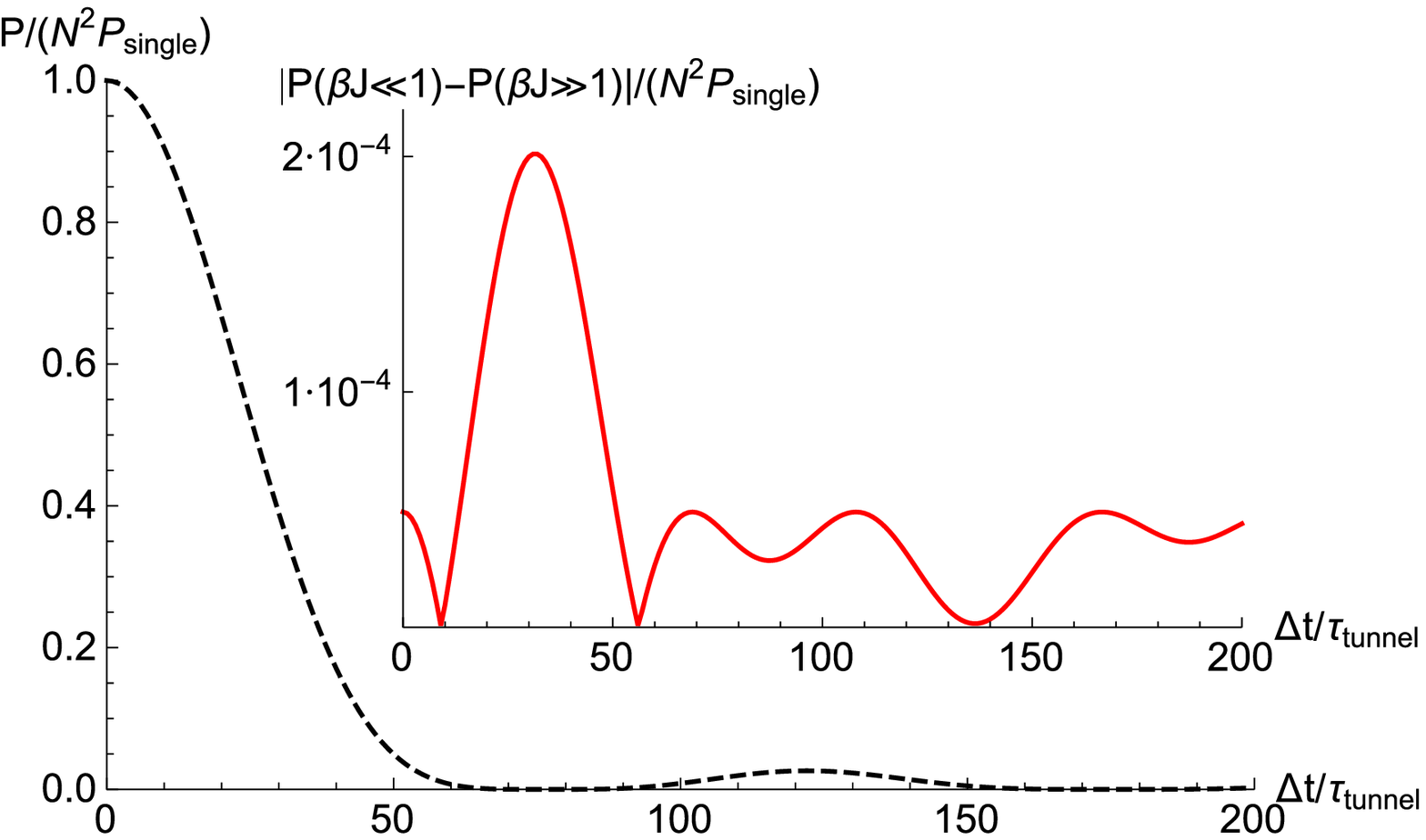}
\caption{
Emission probability $P$ from equation~(\ref{eq:ProbDensHigh})
(normalized by $N^2 P_{\rm single}$, where one factor $N$ originates
from single-photon superradiance and another factor $N$ stems
from the absorption process) for emission
in direction $\Kout=\Kin=2\pi / (L \ell)\left\{1,1\right\}$ 
over the waiting time $\Delta t$ in units of the tunneling time 
$\tau_{\rm tunnel}=\hbar/J$ for $N=L^2=100^2$ 
lattice sites and fermions.
The dashed black line represents the numerical results
for the metallic ground state~(\ref{eq:probabilityFerm}).
For small $\f{\kappa}$, the decay of the superradiance 
peak is identical to the bosonic distributed case 
(lower dashed line in Fig.~\ref{fig:decay_bos}).
As the Fermi-Dirac distribution only interpolates
between these two states with the same emissive behavior,
the decay is similar for arbitrary temperatures
$\beta J$ -- 
aside from the minuscule deviations 
shown in the inset (solid red line).
}
\label{fig:decay_ferm}
\end{center}
\end{figure}

On the other hand,
fermionic ultracold atoms in optical lattices are approximately 
described by the Fermi-Hubbard Hamiltonian~\cite{Hubbard:1963aa,Jaksch:2005kq}.
The two-species Fermi-Hub\-bard Hamiltonian 
can be obtained from the general lattice Hamiltonian (\ref{eq:GenHam})
by defining the possible spin states
$s\in\left\{\uparrow,\downarrow\right\}$, and simplifying
the on-site repulsion term for fermions,
\begin{align}
\label{eq:FHHamExt}
\hat{H}_{\rm FH}^{\rm 2sp}&= -\frac{J}{Z}\sum_{\mu\nu,s,\lambda}T_{\mu\nu}\hat{c}_{\mu s}^{\lambda\,\dagger} \hat{c}_{\nu s}^\lambda\nonumber\\
&\hspace{0.4cm}+ U \sum_{\mu}\left( \sum_{\lambda} \hat{n}_{\mu\uparrow}^{\lambda} \hat{n}_{\mu\downarrow}^{\lambda} + \sum_{\se\sz}\hat{n}_{\mu \se}^{{\rm gr}} \hat{n}_{\mu\sz}^{{\rm ex}}\right)
\,.
\end{align}
Of course, $\hat{c}_{\mu s}^{\lambda\,\dagger}$ and $\hat{c}_{\nu s}^\lambda$
now denote \emph{fermionic} creation and annihilation operators.
To identify plausible initial states of the
fermionic lattice, we explore the ground states of
the single-species Fermi-Hubbard model, which
describes the lattice dynamics prior to the absorption
of the probe photon (i.e., index $\lambda$ omitted),
\begin{align}
\label{eq:FHHam}
\hat{H}_{\rm FH}
=
-\frac{J}{Z}\sum_{\mu\nu,s}\tmn\ccre{\mu s}\cann{\nu s}
+U\sum_{\mu}\hat{n}_{\mu\uparrow} \hat{n}_{\mu\downarrow}
\,.
\end{align}
In the fermionic lattice, we consider the case
of half-filling, with half of the atoms in the $s=\,\uparrow$-state
and half of the atoms in the $s=\,\downarrow$-state,
i.e., $\langle \hat{n}_{\mu\uparrow} \rangle 
= \langle \hat{n}_{\mu\downarrow} \rangle = 1/2$.
Due to the Pauli exclusion principle, the analog of the superfluid state
in the bosonic case is the metallic phase of the Fermi-Hubbard model. 
\setcounter{paragraph}{0}
\paragraph{Mott-N\'{e}el state}
In order to minimize the interaction energy, 
the ground state for $U\gg J$ has to be 
a state with one particle per lattice site,
as in the bosonic case. 
Nevertheless, the ground state is
highly degenerate for $J=0$, as
the energy is unchanged under local spin
rotations.
We will work with the Mott-N\'{e}el state,
which is the approximate ground state
for small but nonzero tunneling rate $J$, 
\begin{align}
\label{eq:NeelFerm}
\ket{\Psi}_{\nee}^{J=0}=\prod\limits_{\mu \in \mathcal{A}} \ccre{\mu\uparrow} \prod\limits_{\nu \in \mathcal{B}} \ccre{\nu\downarrow}\ket{0}=
\bigotimes\limits_{\mu \in \mathcal{A}}\ket{\uparrow}_\mu \bigotimes\limits_{\nu \in \mathcal{B}} \ket{\downarrow}_\nu
\,.
\end{align}
To describe this state, we assume a bipartite lattice, 
i.e., we divide the total lattice into two sub-lattices 
$\mathcal{A}$ and $\mathcal{B}$, 
where for each site $\mu \in \mathcal{A}$,
all the neighboring sites $\nu$ belong to $\mathcal{B}$,
and vice-versa.
All the fermions at the lattice sites in $\mathcal{A}$
have spin $s =\,\uparrow$ while all the fermions at 
the sites in $\mathcal{B}$ have spin $s =\,\downarrow$,
such that neighboring fermions always have opposite
spin (checkerboard pattern).
However, according to Section~\ref{sec:SeparableState},
we obtain the usual single-photon superradiance 
for all states with one atom per lattice site, irregardless of the spin.
In other words, we get the same result for a fermionic lattice 
in the Mott-N\'{e}el state,
as for the Mott insulator state in the case of a bosonic lattice.
\paragraph{Metallic ground state}
For the opposing case of $J \gg U$ and weak interactions, 
the particles arrange in the metallic ground state,
which can be displayed in $\f{k}$-space (Appendix~\ref{ap:kBasis}),
\begin{align}
\label{eq:MetalFerm}
\ket{\Psi}_{\mtl}^{U=0}
=
\prod\limits_{E_{\fk{k}}<E_F,s} \hat c_{\fk{k},s}^\dagger
\ket{0}
\,,
\end{align}
where the product goes over all $\f{k}$-modes whose
energy eigenvalues $E_{\fk{k}}$ are smaller than the Fermi energy $E_F$, 
and all spin orientations
$s\in\left\{\uparrow,\downarrow\right\}$.
In our configuration of a fermionic 2D-lattice with 
half-filling and the dispersion relation of the
Fermi-Hubbard model~(\ref{eq:HJextk}),
$E_{\fk{k}}=-J/Z\, T_{\fk{k}}$ with
$T_{\fk{k}}=2\left[\cos(k_x \ell) + \cos(k_y \ell)\right]$
(for $U=0$),
the occupied modes form
a diamond in two-dimensional $\f{k}$-space whose
vertices touch the borders of the first Brillouin zone,
i.e., a ($\pi/\ell \times \pi/\ell$) square rotated by $\pi/4$%
~\cite{Halboth:1997aa}.
Accordingly, the distribution function in 
Section~\ref{sec:WeakInteractions} is approximately%
\footnote{
In a finite lattice with $N=L^2$ lattice sites (where $L$ is even) 
and exactly $N$ (fermionic) atoms, the metallic ground state
is degenerate, as the edge of the diamond has $4(L-1)$ 
$\f{k}$-modes with the same energy eigenvalue $E_{\fk{k}}$, 
but there are only $2(L-1)$ atoms left to fill these modes.
Thus, we use the distribution function~(\ref{eq:nMetalFerm}),
describing the exact metallic ground state for a modified 
number of $N-2(L-1)\approx N$ atoms, as a (for $L\gg1$)
very good approximation.
} given by:
\begin{align}
\label{eq:nMetalFerm}
n_s^{\mtl}(\f{k}) =
\begin{cases}
1\,, & {\rm if}\,\,|k_x|+|k_y| < \pi/\ell\,,\\
0\,, & {\rm otherwise}\,.
\end{cases}
\end{align}
We obtain the emission probability for the metallic ground state 
by inserting $n_s^{\mtl}(\f{k})$ into equation%
~(\ref{eq:FermionicCorr}) and further into equations%
~(\ref{eq:OpE}) and (\ref{eq:ProbDensHigh}).
It turns out (see Fig.~\ref{fig:decay_ferm}) that a decay of the 
(single-photon) superradiance
peak similar to the bosonic distributed case
(\ref{eq:probability} with $N_1=0$ and $N_2=N$)
occurs (recall the lower dashed line in Fig.~\ref{fig:decay_bos}).
The similarity is quite intuitive, since in both cases
many different $\f{k}$-modes are involved
in the absorption and re-emission of the probe
photon -- all $\f{k}$-modes in the bosonic case
and half of them (the diamond) in the fermionic
case -- instead of just one mode, as in the 
bosonic superfluid ground state.
For small wave numbers $|\f{\kappa}|\ell\ll1$, 
as we suppose in the numerical and experimental
examples, it can be shown analytically 
that the fermionic metallic ground state%
~(\ref{eq:MetalFerm}) gives the same emission probability,
\begin{align}
\label{eq:probabilityFerm}
P=N^2 \left|{\mathcal J}(\Delta t)\right|^2
\delta_{\kin\kout}P_{\rm single}
\,,
\end{align}
as the bosonic distributed case, i.e.,
with the sum over all $\f{k}$-modes,
which can be approximated via Bessel
functions~(\ref{eq:bessels}).
The reason can be found in the phase-sum
as in Appendix~\ref{ap:approx_phasesum}.
Imagine shifting the diamond by e.g.\
$\Delta \f{k} = \pi/\ell \{1,1\}$,
i.e., into the empty regions of the first Brillouin zone.
Doing so, only the sign of the sine-function in equation%
~(\ref{eq:ap:Jxy}) changes, which is irrelevant with 
respect to the symmetric summation.
\paragraph{Thermal state}
For nonzero temperatures, one can 
insert the Fermi-Dirac distribution,
\begin{align}
\label{eq:FDdist}
n^{\rm FD}(\f{k}) = \frac{1}{e^{\beta\left(E_{\fk{k}}-\mu\right)}+1}
\,,
\end{align}
into equation~(\ref{eq:FermionicCorr}).
Numerical results for the emission probability for arbitrary 
temperatures $\beta J$ are hardly different
from the result for the metallic ground state, however
(i.e., the deviations are always smaller than
shown in the inset in Fig.~\ref{fig:decay_ferm}). 
This becomes clear when considering that the 
Fermi-Dirac distribution does nothing else than
to interpolate between the metallic ground state
(low temperature, $\beta J \gg 1$),
and the totally distributed case (high temperature, $\beta J \ll 1$),
where each $\f{k}$-mode is populated by one atom.
As we have just shown that (for small $\f{\kappa}$)
these both extreme ends of the temperature scale 
show the same analytic result~(\ref{eq:probabilityFerm}),
it is clear that interpolating between these
two extremes does not produce qualitatively
different results.

In conclusion for a fermionic lattice, we obtained 
the usual single-photon superradiance only 
for the Mott-N\'{e}el state (Sect.~\ref{sec:SeparableState}).
The metallic ground state as well as (partially) excited states 
show a reduction in superradiance.
Thus, as opposed to the bosonic case, fermionic
(partially) excited states cannot be distinguished from
the metallic ground state in the weak interactions regime.
However, in the fermionic case, the Mott-N\'{e}el state
can be distinguished from the metallic ground state,
i.e., the current parameter regime ($J\gg U$ or $U\gg J$)
can be detected using our probe.
%
\section{Phase transitions}
\label{sec:Transitions}
%
In the previous sections, we discussed to what extend the proposed
probe can be used to (nondestructively) detect
certain lattice states in either of the limiting cases.
As another application, we will now discuss
the probe of a quantum phase transition from the
separable state regime (small $J$) to the 
weak interactions regime ($J \gg U$).
To this end, we consider a similar probing sequence 
as in Figure~\ref{fig:probe_dicke}.
First, the probe photon $\Kin$ is absorbed by 
the bosonic or fermionic lattice, which is initially in the 
Mott insulator state~(\ref{eq:MottBos}) 
or the Mott-N\'{e}el state~(\ref{eq:NeelFerm}), respectively.
Afterwards, either a sudden quantum quench or an adiabatic
transition to the weak interactions regime is applied.
Then, the sequence continues as before -- after
a waiting period $\Delta t$, in which the atoms
have time to tunnel and interact,
the probe photon is re-emitted.

Analogous to equation~(\ref{eq:ProbDens1}),
we consider the absorption of the probe photon
in first order perturbation theory, leading to the
(non-normalized) excited state
\begin{align}
\label{eq:MottNeelEx}
\ket{\Psi}_{\exc}^{\mon}
=
-i \int_0^{\tau_A}\hspace{-0.2cm}dt_1\,\hat{V}_U(t_1)
\ket{\Psi}_{\mon}^{J=0}\hat{a}_{\kin}^\dagger\ket{0}
\,,
\end{align}
where the index $U$ of the perturbation Hamiltonian denotes
that only interaction applies, as we are in the separable state
regime ($J=0$) in the beginning.
In addition, the superscript $\mot$ indicates the bosonic Mott insulator 
state~(\ref{eq:MottBos}), while $\nee$ is an abbreviation for the 
fermionic Mott-N\'{e}el state~(\ref{eq:NeelFerm}). 
As suggested by the diagonal slash, we consider both
cases simultaneously where possible.

\subsection{Sudden quantum quench}
In the case of a sudden quantum quench, the initial excited state%
~(\ref{eq:MottNeelEx}) cannot adapt adiabatically and is 
thus not an eigenstate of the new ($U=0$) Hamiltonian.
We now calculate the emission probability
density after such a sudden transition, i.e., starting
from the excited Mott insulator state or 
the excited Mott-N\'{e}el state~(\ref{eq:MottNeelEx}).
Applying first order perturbation theory for the 
emission process (the index $J$ denotes that
only tunneling applies, as we are in the weak
interactions regime, $U=0$, after the transition),
\begin{align}
\label{eq:PMottexcited}
P
=\left\|\bra{0}\hat{a}_{\kout}\int_0^{\tau_E}dt_2\,\hat{V}_J(t_2)
\ket{\Psi}_{\exc}^{\mon}\right\|^2
\,,
\end{align}
we obtain an expression analog to equation~(\ref{eq:ProbDens2}).
For convenience, we insert both the exciton creation operator
and the exciton annihilation operator in the $\f{k}$-basis
(as in Sect.~\ref{sec:WeakInteractions}).
However, as we are still in the separable state ($J=0$) regime
during exciton creation $\hat{\Sigma}_U^+(\Kin,t_1)$,
its phase factor~(\ref{eq:InterPhase}) does not emerge here,
\begin{align}
&\hat{\Sigma}_J^-\left(\Kout,t_2\right)
\hat{\Sigma}_U^+(\Kin,t_1)\ket{\Psi}_{\mon}^{J=0}
=e^{i\omega (t_1-t_2)}\\
&\times
\sum_{\fk{k},s} \exp\left\{-i\phi_{\fk{k}}^{\kout}(t_2)\right\}
\hat{a}_{\fk{k}-\kout,s}^{{\rm gr}\,\dagger}\hat{a}_{\fk{k}-\kin,s}^{{\rm gr}}\ket{\Psi}_{\mon}^{J=0}\nonumber
\,.
\end{align}
Accordingly, the same four-point correlator 
as in equation~(\ref{eq:OpE}) has to be calculated --
but now the expectation value has to be taken
in the Mott state~(\ref{eq:MottBos}) or the
Mott-N\'{e}el state~(\ref{eq:NeelFerm}),
respectively.
The bosonic correlator as in equation~(\ref{eq:BosonicCorr}),
taken in the Mott insulator state $\ket{\Psi}_{\mot}^{J=0}$,
yields
\begin{align}
E_{\mot}^{\rm B} =
\delta_{\kin \kout} + 2 \delta_{\fk{k} \fk{q}} - 2/N
\,,
\end{align}
while the fermionic correlator as in~(\ref{eq:FermionicCorr}),
taken in the Mott-N\'{e}el state $\ket{\Psi}_{\nee}^{J=0}$, gives
\begin{align}
\sum_{\se\sz}
E_{\nee}^{\rm F} =
\delta_{\kin\kout} +
\delta_{\fk{k}\fk{q}}/2
\,.
\end{align}
For details on the calculation, see Appendix~\ref{ap:FourPointCorrMott}.
As in the derivation in Appendix~\ref{ap:PartialCond}, we only keep the
terms which contribute quadratically in $N$ to the emission probability
density~(\ref{eq:PMottexcited}), i.e., the leading order terms which 
correspond to coherent (single-photon) superradiant emission, as indicated
by the factor $\delta_{\kin\kout}$.
The resulting emission probability density~(\ref{eq:PMottexcited}) in
the same direction in which the absorbed photon was incoming,
$\Ka = \Kin = \Kout$, is thus given by
\begin{align}
P
=
N^2 \delta_{\kin\kout}\,\mathcal{I}\left[ e^{i(\omega_{\rm out}-\omega)t_2}\mathcal{J}(t_2) \right]\,.
\end{align}
So far, we did not account for the waiting time $\Delta t$.
Here, we model it via the time-dependent coupling constant
$g_{\kout}^*(t_2)$ in equation~(\ref{eq:ProbDensFunc}), 
which should be zero for a period $\Delta t$ after the 
sudden transition and before the emission process 
starts as usual, e.g.,
\bea
g_{\kout}^*(t_2)
= 
\begin{cases}
0,\quad & \hspace{-0.5cm}\mbox{if }\quad t_2<\Delta t\,,\\
g_{\kout}^*,\quad & \hspace{-0.5cm}\mbox{if }\quad t_2\geq\Delta t
\,.
\end{cases}
\ea
As the time span of the emission process is negligible
in comparison to the waiting time $\Delta t$, we can 
approximate $\mathcal{J}(t_2)=\mathcal{J}(\Delta t)$
and thus arrive at
\begin{align}
\label{eq:PsuddenBos}
P
&=
N^2 \left|\mathcal{J}(\Delta t)\right|^2 \delta_{\kin\kout} P_{\rm single}
\,.
\end{align}
That is, the emission probability after a sudden switching procedure
shows a reduction in superradiance similar to the bosonic
distributed case~(\ref{eq:probability} with $N_1=0$ and $N_2=N$)
and the fermionic metallic ground state~(\ref{eq:probabilityFerm}).
The result is quite intuitive -- after the sudden switch,
the Mott (or Mott-N\'{e}el) state behaves as a state where 
all momenta are equally populated, as it appears as a 
mixture of excited states to the new ($U=0$) Hamiltonian.

\subsection{Adiabatic transition}
In contrast to a quantum quench, i.e., an abrupt change of the parameter 
regime, let us now consider an adiabatic transition, i.e., a very slow 
change of the parameter in a finite lattice -- such that the system stays 
close to its instantaneous ground (or excited) state. 
The case of a bosonic lattice was already studied in reference%
~\cite{Brinke:2015aa}, where we came to the conclusion that the 
superradiance peak does not decay after an adiabatic transition, 
i.e., the superradiant emission probability reads 
$P = N^2 \delta_{\kin\kout} P_{\rm single}$. For a detailed 
derivation of this result, see Appendix~\ref{ap:AdiabTransBos}.

Summing up for a bosonic lattice, the (single-photon)
superradiance peak thus decays according to equation%
~(\ref{eq:PsuddenBos}) after a sudden quantum quench, 
while it does not decay in the case of an adiabatic transition.
Hence, employing the proposed probing scheme, a sudden transition 
can be distinguished from an adiabatic transition via the different
emission characteristics.

Let us now study the fermionic case, which will
yield a different result.
In principal, starting from equation~(\ref{eq:MottNeelEx}),
the same reasoning as in the bosonic case applies.
Anyway, in the case of fermions, we cannot 
unambiguously identify the corresponding eigenstate
in the weak interactions regime.
Taking, e.g., the metallic ground state~(\ref{eq:MetalFerm}) 
as a reasonable starting point, it is not clear which of the
various $\f{k}$-modes would carry the exciton wave number 
$\Kin$ --
whereas in the bosonic case there is obviously only one 
possibility.
However, a first approximation for small wave vectors $\f{\kappa}$
can be obtained by considering the commutator between
the (time-dependent) Hamiltonian $\hat{H}_{\rm A}(t)$ modeling 
the adiabatic evolution and the exciton creation operator, 
$\hat{\Sigma}_U^+(\Kin,t_1)$,
\begin{multline}
\left[ \hat{H}_{\rm A}(t), \hat{\Sigma}_U^+(\Kin,t_1) \right]
=
-\frac{J(t)}{Z}e^{i\omega t_1}\\
\times\sum_{\fk{p},s}\hat{c}_{\fk{p}-\kin,s}^{{\rm gr}} \hat{c}_{\fk{p},s}^{{\rm ex}\,\dagger} \left( T_{\fk{p}} - T_{\fk{p}-\kin} \right)
\,.
\end{multline}
Notably, the commutator scales linearly in $(\Kin)_{x/y}$ for 
small wave vectors $|\Kin|\ell\ll1$, see~(\ref{eq:ap:ApproxSmallKappa})
in Appendix~\ref{ap:approx_phasesum}.
Presuming that the commutator is negligible,
we can apply the adiabatic evolution 
$\hat{U}_A(t) \ket{\Psi}_{\nee}^{J=0} = \ket{\Psi}_{\mtl}^{U=0}$
before the exciton creation operator instead of afterwards.
Then, however, the emission probability~(\ref{eq:PMottexcited})
can be calculated via
\begin{multline}
\hat{\Sigma}_J^-\left(\Kout,t_2\right)
\hat{\Sigma}_U^+(\Kin,t_1)\ket{\Psi}_{\mtl}^{U=0}
=e^{i\omega (t_1-t_2)}\\
\times
\sum_{\fk{k},s} \exp\left\{-i\phi_{\fk{k}}^{\kout}(t_2)\right\}
\hat{c}_{\fk{k}-\kout,s}^{{\rm gr}\,\dagger}\hat{c}_{\fk{k}-\kin,s}^{{\rm gr}}\ket{\Psi}_{\mtl}^{U=0}
\,,
\end{multline}
which traces back to the case of the metallic ground state,
for which we already showed that the superradiance peak decays as%
~(\ref{eq:probabilityFerm}), i.e.,
\begin{align}
\label{eq:PadiabFerm}
P=N^2 \left|{\mathcal J}(\Delta t)\right|^2
\delta_{\kin\kout}P_{\rm single}
\,.
\end{align}
Even in the general case, where the adiabatic evolution
does not commute with exciton creation,
the quantum state after the adiabatic evolution
has to be found somewhere in the spectrum between
the metallic ground state and a thermal state.
As we have shown in Section~\ref{sec:FermiHubbard}
and is particularly clear in Figure~\ref{fig:decay_ferm}, 
both ends of this spectrum show the same reduction in 
superradiance~(\ref{eq:PadiabFerm}).
So in conclusion for a fermionic lattice, 
we expect a decay of the superradiance peak after 
a sudden quantum quench as well as after an adiabatic evolution.
Thus, unfortunately, and in contrast to the phase transition in the 
bosonic lattice, we cannot distinguish a sudden transition
from an adiabatic transition in a fermionic lattice.
%
\section{Experimental realization}
\label{sec:Experimental}
%
After having developed the theoretical framework
of our probing scheme both for probing the quantum
state of an optical lattice (Sect.~\ref{sec:ProbingLatticeStates}), 
as well as properties of phase transitions (Sect.~\ref{sec:Transitions}), 
we want to discuss the overall parameters for experimental implementation.
As mentioned in Section~\ref{sec:model}, it is desirable that
the probe-photon wavelength $\lambda_{\rm photon}$ is large compared 
to the lattice spacing $\ell=\lambda_{\rm lat}/2$.
Thus, ideally, one would like to have a small lattice spacing, e.g.,
a green lattice $\ell=257\;{\rm nm}$%
~\cite{Inouye:1998fk,Jaksch:1998uq,Jaksch:2005kq} created by 
an argon-ion~\cite{Bridges:1964ly} laser with $\lambda_{\rm lat}=
514\;{\rm nm}$.
On the other side, using an infrared probe photon with a wavelength of
$\lambda_{\rm photon}=2\pi/|\Kin|=10.6\;\mu{\rm m}$~\cite{Patel:1964fk},
then makes sure that the collective coherent emission (i.e., Dicke superradiance) outpaces spontaneous incoherent emission processes by a wide margin~\cite{Brinke:2013fk}.
Moreover, the atomic recoil is negligible in case of an infrared photon,
as its recoil energy is a factor $E_R^{\rm lat}/E_R^{\rm photon} = 4\cdot10^2$ 
smaller than the recoil energy of an optical lattice photon.
This specific combination of trapping atoms (or molecules) which 
feature a far-infrared transition in a green lattice may be difficult to 
achieve experimentally. 
However, recent calculations and experiments~\cite{Bettles:2015aa,
Bromley:2016aa} suggest that cooperative effects still dominate even 
when the lattice spacing $\ell$ is only slightly smaller (but still of the 
same order) than the driving wavelength $\lambda_{\rm photon}$.
Hence, on the one hand, it should also be feasible to use larger optical
lattice wavelengths, e.g., in the common range $\lambda_{\rm lat}=
500-1000\;{\rm nm}$~\cite{Stamper-Kurn:1998uq, Greiner:2002fk, 
Kohl:2005aa,Volz:2006aa,Jordens:2008aa,Moses:2015aa}, which is 
beneficial as the lattice wavelength can then be chosen to fit the 
properties of the trapped atoms.
On the other hand, a smaller optical wavelength, e.g., in the 
near-infrared, $\lambda_{\rm photon}=1-10\;\mu{\rm m}$,
should also be possible.
As for the required number of atoms, we envisage it in the typical range
of 2D optical lattice experiments~\cite{Spielman:2007kx,Gemelke:2009aa,Sherson:2010uq,Endres:2013fk,Greif:2016aa}, i.e., in the range of 
$N=10^2-10^4$ atoms and lattice sites. 
Of course, the higher the number of involved atoms, the
stronger is the utilized effect of single-photon superradiance which leads
to an increased sensitivity of the proposed probe. On the other hand, it 
becomes harder to fulfill the assumption of equal coupling~(\ref{eq:hInt}).
Below, we reflect on four possible setups for experimental
realization (including the next Sect.~\ref{sec:Classical}).

\subsection{Two-level system}
The probing sequence displayed in Figure~\ref{fig:probe_dicke}
basically poses two requirements on the physical system.
First and foremost, the atoms in the optical lattice need
to support an infrared transition via their level scheme,
in order to be able to absorb and re-emit the infrared probe photon.
Second, the lifetime of this transition needs to be long as
compared to the time scales of the lattices dynamics, i.e.,
(much) longer than the typical tunneling time of
$\tau_{\rm tunnel} = \hbar/J = 5\cdot10^{-5}\;{\rm s}$,
such that significant tunneling can occur before the re-emission.
Hence, two-level atoms with a corresponding level scheme
would suffice to implement the probing sequence.
Although in principle the scheme could be implemented via
a simple two-level system,
the requirements are probably hard to meet when focusing on
the atoms typically used in optical lattice experiments (e.g., Rb, Na).
However, choosing different atoms (elements) or even considering 
molecules~\cite{Volz:2006aa,Moses:2015aa} may render a direct 
implementation possible.

\subsection{Assisted multi-photon transition}
\begin{figure}[t]
\begin{center}
\subfigure[]{\includegraphics[width=0.4\columnwidth]{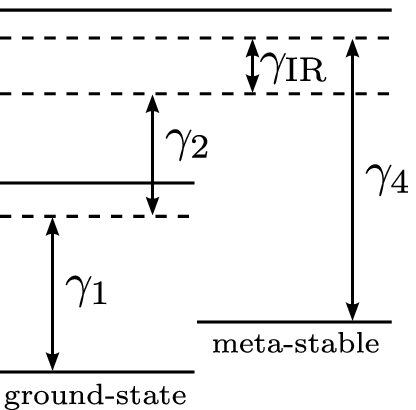}}
\hspace{.5cm}
\subfigure[]{\includegraphics[width=0.4\columnwidth]{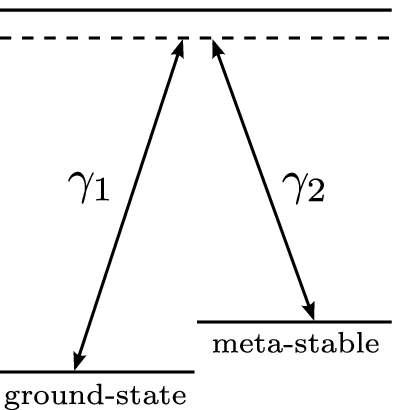}}
\caption{Two examples of level schemes proposed for
the experimental implementation. (a) In a detuned four-photon transition,
the infrared probe photon $\gamma_{\rm IR}$ can only be absorbed
or emitted under assistance of the laser fields $\gamma_1$, 
$\gamma_2$, and $\gamma_4$.
(b) Detuned two-photon Raman transition. Excitation either via
a single photon $\gamma_1$ (where a Stokes photon $\gamma_2$
is scattered), or via two counter-propagating lasers $\gamma_1$
and $\gamma_2$.
}
\label{fig:infrared_and_classical}
\end{center}
\end{figure}

In order to overcome the requirement on the transition lifetime
and, in addition, gain full experimental control over the
waiting time $\Delta t$, one can replace the single-photon
transition envisaged above by a laser-assisted 
multi-photon transition.
In this scenario, the infrared transition of the probe photon
is enabled only while the assisting lasers are switched on.
Thus, switching the lasers on for the absorption and 
emission of the infrared probe photon, but switching
them off for an arbitrary waiting time $\Delta t$ in between,
leaves the control of the involved time scales at the
discretion of the experimentalist. 
Note that then also long waiting times of, e.g., 
$\Delta t = \ord(10^2)\cdot \tau_{\rm tunnel}$
are feasible, preferable to best distinguish 
different states or phase transitions
(see, e.g., Fig.~\ref{fig:decay_bos}).
An example of a detuned four-photon transition
is presented in Figure~\ref{fig:infrared_and_classical}a.
Here, the four-photon transition involving the 
absorption or emission of an infrared probe photon $\gamma_{\rm IR}$
is assisted by three photons $\gamma_1$, $\gamma_2$, and $\gamma_4$,
which are provided by the assisting external lasers.
Note that the four-photon transition as depicted in 
Figure~\ref{fig:infrared_and_classical}a only serves as an 
example -- the assisting lasers can be chosen as to
match a wide variety of given level schemes.
\subsection{Two-photon Raman process}
One can also think of a two-photon Stokes
Raman scattering process to create the Dicke state among
the optical lattice atoms, see, e.g.,~\cite{Scully:2006fk,Scully:2007fk}.
For example, consider the level scheme shown in
Figure~\ref{fig:infrared_and_classical}b.
An incident photon $\gamma_1$ leads to an excitation
of the metastable level via the scattering of a Stokes
photon $\gamma_2$ with wave vector $\f{k}_2$.
The excitation vector $\Kin$ of the created Dicke state
can then be determined by detecting the Stokes photon,
i.e., $\Kin = \f{k}_1 - \f{k}_2$, and should be in the infrared
region.
Analogously, the emission process would correspond to
an anti-Stokes Raman scattering, 
initiated by a suitable pulse on demand.
%
\section{Classical laser fields}
\label{sec:Classical}
%
As a fourth option, we would like to present an 
alternative probing scheme, employing coherent states 
as generated by classical laser fields, instead
of single-photon absorption and emission.
While working with classical laser fields may be experimentally
less challenging than the absorption and emission of single photons, 
we will show that main results of our probing scheme are retained,
although the excitation (and deexcitation) dynamics differ.
Basically, we again consider the two-photon (Raman) 
transition as given in Figure~\ref{fig:infrared_and_classical}b,
but this time operated via two counter-propagating lasers 
$\gamma_1$ and $\gamma_2$.
We envisage the following modified sequence. First, 
two counter-propagating lasers ($\gamma_1$ and $\gamma_2$)
are switched on for a short time $t_{\rm in}$ to excite 
lattice atoms from the ground state to a metastable state
via the detuned two-photon (Raman) transition depicted
in Figure~\ref{fig:infrared_and_classical}b.
Then, during a waiting time $\Delta t$ in which the lasers are 
turned off, the atoms tunnel and interact according to
equation~(\ref{eq:GenHam}).
Finally, the two counter-propagating lasers are turned on 
for a second time $t_{\rm out}$, stimulating the decay 
from the metastable state back to the ground state.

Instead of the perturbation Hamiltonian~(\ref{eq:hInt})
for the free-space electromagnetic field, we treat the
electromagnetic field classically via 
\begin{align}
\label{eq:hIntClassical}
\hat{V}_{\rm cl} = g A_{\rm eff}(t)\hat{\Sigma}^+\left(\f{\kappa}\right) 
+ {\rm H.c.}
\,,
\end{align}
where the atoms in the optical lattice effectively
interact with only one mode $\f{\kappa}$, which is given
by the momentum difference $\f{\kappa} = \f{k}_1 - \f{k}_2$ 
of the counter-propagating lasers.
Furthermore, $A_{\rm eff}(t)$ describes the
effective classical field of the two counter-propagating
lasers, and $g$ denotes an effective coupling constant
depending on atomic properties.
To simplify the upcoming calculations, we assume
it to be real, which leads to
\begin{align}
\hat{V}_{\rm cl} = g A_{\rm eff}(t)\left[ \hat{\Sigma}^+\left(\f{\kappa}\right) + {\rm H.c.} \right]
= 2 g A_{\rm eff}(t)\hat{\Sigma}^x\left(\f{\kappa}\right)
\,.
\end{align}
Of course, $\hat{\Sigma}^x\left(\f{\kappa}\right)$ is the generator
of a quasispin-rotation about the $x$-axis. 
Hence, during the excitation period $t_{\rm in}$, 
the quasispin-$Z$-operator gets rotated 
by an angle $\alpha = 2 g \int_{0}^{t_{\rm in} }dt\,A_{\rm eff}(t)$, i.e.,
\begin{align}
\label{eq:SigmaZExc}
\hat{\Sigma}^z_{\alpha}&=
\hat{U}_{\rm cl}^\dagger(\alpha)\hat{\Sigma}^z\hat{U}_{\rm cl}(\alpha)\nonumber\\
&=\cos\left(\alpha\right) \hat{\Sigma}^z + \sin\left(\alpha\right) \hat{\Sigma}^y\left(\f{\kappa}\right)
\,.
\end{align}
Afterwards, the atoms in the optical lattice 
can tunnel according to the general lattice
Hamiltonian~(\ref{eq:GenHam}).
In the weak interactions regime ($U=0$), the
subsequent tunneling~(\ref{eq:HJextk}) over the waiting 
period $\Delta t$ can be expressed in the $\f{k}$-basis 
via the time-evolution operator
\begin{align}
\hat{U}_J\left(\Delta t\right) 
= 
\exp\Bigg\{i \frac{J \Delta t}{Z}\sum_{\fk{k},s,\lambda}T_{\fk{k}}^{\phantom{\,}}
\hat{n}_{\fk{k},s}^{\lambda}\Bigg\}
\,.
\end{align}
Applying the tunneling dynamics to the rotated quasispin-$Z$-operator, which 
already contains the excitation process~(\ref{eq:SigmaZExc}), yields
\begin{align}
\label{eq:SigmaZExcTun}
\hat{\Sigma}^z_{\alpha,\Delta t} &= \hat{U}_J^\dagger\left(\Delta t\right) \hat{\Sigma}^z_{\alpha} \hat{U}_J\left(\Delta t\right)\nonumber\\
&= \cos\left(\alpha\right) \hat{\Sigma}^z + \sin\left(\alpha\right) \hat{\Sigma}^y_{J}\left(\f{\kappa}\right)
\,,
\end{align}
where $\hat{U}_J^\dagger\left(\Delta t\right) \hat{\Sigma}^z\hat{U}_J\left(\Delta t\right)$ 
is unchanged (only number operators are involved), while 
$\hat{\Sigma}^y\left(\f{\kappa}\right)$ picks up mode-specific
phases~(\ref{eq:InterPhase})
similar to the interaction picture considerations
above~(\ref{eq:ExCreIntPic}),
\begin{multline}
\hat{\Sigma}^y_J\left(\f{\kappa}\right) := \hat{U}_J^\dagger\left(\Delta t\right)\hat{\Sigma}^y\left(\f{\kappa}\right)\hat{U}_J\left(\Delta t\right)\\
=\frac{1}{2i}\Bigg(\sum_{\fk{p},s} 
\hat{a}_{\fk{p},s}^{{\rm ex}\,\dagger}\hat{a}_{\fk{p}-\fk{\kappa},s}^{{\rm gr}} \exp\{i \phi_{\fk{p}}^{\fk{\kappa}}(\Delta t)\} - {\rm H.c.}\Bigg)
\,.
\end{multline}
Then, in the third and final step, the counter-propagating lasers 
are switched on for a second time period $t_{\rm out}$ in order
to reverse the rotation $\alpha$.
Hence, we apply the appropriate time-evolution $\hat{U}_{\rm cl}(\beta)$ 
with $\beta = 2 g \int_{0}^{t_{\rm out} }dt\,A_{\rm eff}(t)$
to equation~(\ref{eq:SigmaZExcTun}) to complete the sequence.
After some lengthy but straightforward (commutator) 
calculations we arrive at:
\begin{align}
\label{eq:SigmaZResult}
\hat{\Sigma}^z_{\alpha,\Delta t,\beta} 
&= \hat{U}_{\rm cl}^\dagger(\beta)\hat{\Sigma}^z_{\alpha,\Delta t}\hat{U}_{\rm cl}(\beta)\nonumber\\
&=\cos\left(\alpha\right)\cos\left(\beta\right)\hat{\Sigma}^z
+ \cos\left(\alpha\right)\sin\left(\beta\right)\hat{\Sigma}^y\left(\f{\kappa}\right)\nonumber\\
&\hspace{0.4cm}-\sin\left(\alpha\right)\sin\left(\beta\right)\hat{\Sigma}^z_C\left(\f{\kappa}\right)+ \sin\left(\alpha\right)\hat{\Sigma}^y_{J}\left(\f{\kappa}\right)\nonumber\\
&\hspace{0.4cm}+\sin\left(\alpha\right)\left[\cos\left(\beta\right)-1\right] \hat{\Sigma}^y_C\left(\f{\kappa}\right)
\,,
\end{align}
where we defined
\begin{align}
\hat{\Sigma}^z_C\left(\f{\kappa}\right)&:=\frac{1}{2}\sum_{\fk{p},s}\left(\hat{n}_{\fk{p},s}^{{\rm ex}}-\hat{n}_{\fk{p}-\fk{\kappa},s}^{{\rm gr}}\right) \cos\left\{\phi_{\fk{p}}^{\fk{\kappa}}(\Delta t)\right\}\,,\\
\hspace{-1.0cm}\hat{\Sigma}^y_C\left(\f{\kappa}\right)&:=\frac{1}{2i}\sum_{\fk{p},s}\left(\hat{a}_{\fk{p},s}^{{\rm ex}\,\dagger}\hat{a}_{\fk{p}-\fk{\kappa},s}^{{\rm gr}} - {\rm H.c.}\right)\cos\left\{\phi_{\fk{p}}^{\fk{\kappa}}(\Delta t)\right\}
\,.
\end{align}
Now, the full sequence is modeled in equation~(\ref{eq:SigmaZResult}).
Hence, we can evaluate the expectation value 
$\bra{\Psi}\hat{\Sigma}^z_{\alpha,\Delta t,\beta}\ket{\Psi}$
-- which is just the (shifted by $N/2$) number of atoms still excited
after the completed sequence --
for an arbitrary initial state $\ket{\Psi}$.
As before, we only consider initial states $\ket{\Psi}$
which do not contain any excitations. Thus,
all $\hat{\Sigma}^y$-related expectation values vanish,
$\bra{\Psi}\hat{\Sigma}^z\ket{\Psi}=-N/2$ and
\begin{align}
\bra{\Psi}\hat{\Sigma}^z_C\left(\f{\kappa}\right)\ket{\Psi}
=-\frac{1}{2}\sum_{\fk{p},s} n_s(\f{p}-\f{\kappa})\cos\left\{\phi_{\fk{p}}^{\fk{\kappa}}(\Delta t)\right\}
\,,
\end{align}
where $n_s(\f{p}) = \bra{\Psi}\hat{n}_{\fk{p},s}^{{\rm gr}}\ket{\Psi}$.
In summary, we arrive at
\begin{multline}
\label{eq:SigmaZPsi}
\langle\hat{\Sigma}^z_{\alpha,\Delta t,\beta}\rangle
=-\frac{N}{2}\cos\left(\alpha\right)\cos\left(\beta\right)+\frac{1}{2}\sin\left(\alpha\right)\sin\left(\beta\right)\\
\times\sum_{\fk{p},s} n_s(\f{p}-\f{\kappa})\cos\left\{\phi_{\fk{p}}^{\fk{\kappa}}(\Delta t)\right\}
\,.
\end{multline}
To point out the differences and similarities as compared to
the single-photon approach, let us assume very short laser periods
$\alpha\ll1$, $\beta=-\alpha$ 
(where the sign of $A_{\rm eff}(t)$ is reversed
for the second time period $t_{\rm out}$)
and second order approximation in $\alpha$. 
Then, according to~(\ref{eq:SigmaZExc}),
$\overline{n}= \langle\hat{\Sigma}^z\rangle + N/2 \approx N \alpha^2 /4$
atoms are excited in the first step.
If, for example, $\overline{n}$ is smaller than unity, the resulting
coherent state is well approximated by a coherent superposition of the ground state 
$\ket{\rm ground}$ with $\hat{\Sigma}^-\left(\f{\kappa}\right)\ket{\rm ground}=0$ 
and the first excited Dicke state $\hat{\Sigma}^+\left(\f{\kappa}\right)\ket{\rm ground}$, 
which would be the stand-alone result of a single-photon excitation.
In the same approximation, the number of atoms still in the metastable state
after the full sequence is then given by:
\begin{multline}
\langle \hat{n}_{\rm meta} \rangle 
=
\langle\hat{\Sigma}^z_{\alpha,\Delta t,\beta}\rangle + \frac{N}{2}\\
=2\overline{n}\left(1-\frac{1}{N}\sum_{\fk{p},s} n_s(\f{p}-\f{\kappa})\cos\left\{\phi_{\fk{p}}^{\fk{\kappa}}(\Delta t)\right\}\right)
\,,
\end{multline}
according to equation~(\ref{eq:SigmaZPsi}).
Inserting, for example, the bosonic partial condensation state
with $N_1$ atoms condensed at $\f{k}=0$ 
and $N_2$ atoms equally distributed over all $\f{k}$-modes, 
which led to the emission probability~(\ref{eq:probability}), we find
\begin{align}
\label{eq:probability_classical}
\langle \hat{n}_{\rm meta} \rangle 
&=2\overline{n}\left(1-\frac{N_1}{N}\cos\left\{\varphi(\Delta t)\right\}-\frac{N_2}{N}\mathcal{J}(\Delta t)\right)
\,,
\end{align}
where we considered that the phase-sum over cosine equals
the phase-sum over the exponential function~(\ref{eq:reduction})
in Appendix~\ref{ap:PartialCond}, 
as the imaginary part is zero anyway.
Comparing this result to the result of the single-photon approach, 
we see that we can infer the number $N_1$
of condensed atoms in both cases -- 
either via the emission characteristics
in the single-photon approach~(\ref{eq:probability}), 
or by measuring the number of atoms which are still in the 
meta\-stable state after the full sequence~(\ref{eq:probability_classical}).
Going into detail, the term $N_2{\mathcal J}(\Delta t)$ 
represents the deterioration of the spatial phase coherence,
as in the single-photon approach above.
As a consequence, the deexcitation process 
via the two counter-propagating lasers is hindered,
as the altered spatial phases do not fit to the combined
laser wave vector $\f{\kappa} = \f{k}_1 - \f{k}_2$ anymore.
In the quasispin picture, the rotation $\alpha$ about the $x$-axis%
~(\ref{eq:SigmaZExc}) would be perfectly rotated back via the 
second rotation about $\beta=-\alpha$, leading to 
$\langle \hat{n}_{\rm meta} \rangle=0$, if there would
be no hopping in between ($J=0$).
For nonzero tunneling $J>0$, however, the back-rotation
is not perfect, which leads to a finite probability
for excited atoms remaining in the final state%
~(\ref{eq:probability_classical}).
Interestingly, the phase $\varphi(\Delta t)$ occurs in
a cosine here, as compared to an exponential phase
factor in equation~(\ref{eq:probability}).
However, this discrepancy is a direct consequence of the 
classical approach, where we do not create an exact 
Dicke state but rather a coherent superposition of excited states
with different energies.
In the quasispin picture, the cosine-term can be understood
as a kind of Larmor precession of the quasispin vector
about the $z$-axis during the waiting time $\Delta t$.

In conclusion, similar results as in the single-photon
approach can be obtained using classical laser fields,
which may be more feasible experimentally.
%
\section{Conclusions}
\label{sec:Conclusions}
%
In summary, we employed Dicke superradiance, i.e.,
the collective and coherent absorption and emission
of photons, 
to probe the quantum state, or properties of a 
quantum phase transition, of an ensemble of 
ultracold atoms in an optical lattice.
In a detailed analysis, we studied both single-photon
superradiance~(\ref{eq:hInt}) as well as superradiance
in the context of classical laser fields~(\ref{eq:hIntClassical}),
while the lattice dynamics was modeled via a general
lattice Hamiltonian~(\ref{eq:GenHam}), covering both
the Bose-Hubbard model and the Fermi-Hubbard model
as special cases.

We found that the (single-photon) superradiant 
emission characteristics are significantly altered due to 
the lattice dynamics in a variety of scenarios, which can be utilized
to obtain information about (the evolution of) the quantum
state of the atoms.
Considering a bosonic optical lattice in the
weak interactions regime ($J\gg U$), (partially) excited
states, such as e.g.\ thermal states, can be distinguished
from condensed states.
It may even be possible to infer the number of condensed
atoms~(\ref{eq:probability}), 
or the temperature (see Fig~\ref{fig:decay_bos}),
respectively.
In a fermion\-ic optical lattice, our probe can distinguish
the metallic ground state (reduction in super\-radiance)
from the Mott-Néel ground state (no reduction),
i.e., detect the current parameter regime 
($J\gg U$ or $U\gg J$).

Regarding quantum phase transitions, 
it is possible to discriminate between an adiabatic transition 
(no reduction) and a sudden quench transition 
(reduction in superradiance) from the Mott insulator
state to the superfluid phase in a bosonic lattice.
Unfortunately, such a distinction can not be made
in case of a fermionic lattice, as the superradiance peak
decays similarly after either transition (see Fig.~\ref{fig:decay_ferm}).
Finally, we discussed several options for an experimental realization
of the single-photon probing scheme (see Fig.~\ref{fig:probe_dicke}),
as well as a modified probing sequence employing classical laser fields.

We would like to stress again that our method is complementary to other
techniques (e.g., Bragg-scattering), as information about the temporal
evolution of the coherence properties of the atoms is encoded in the
emission characteristics, and thus can be measured nondestructively.
Hence, also non-equilibrium spectral information can be obtained,
analogous to pump-probe spectroscopy in solid-state physics.
As another topic of recent interest, note that we may distinguish
even and odd-frequency correlators~\cite{Berezinskii:1974aa},
as access to double-time Green functions~\cite{Zubarev:1960aa}
is obtained via the different time coordinates included in the
correlator~(\ref{eq:OpD}).
Further research could e.g.\ focus on whether 
superconductivity also shows signatures 
in the correlator~(\ref{eq:OpD}).

\begin{acknowledgement}

The authors would like to thank Konstantin Krutitsky and
Friedemann Queisser for valuable discussions.
This work was supported by the DFG (SFB-TR12). 
\end{acknowledgement}
\appendix
\section{Correlators $\mathcal{C}$ of the operatorial function $\mathcal{D}$}
\label{ap:DCorrelators}
To calculate the operatorial function~(\ref{eq:OpD}) 
for the case of a separable state with small correlations 
between lattice sites, it is necessary to go through all 
possible cases for the correlator~(\ref{eq:Corr1}) 
\begin{multline}
\hspace{-0.2cm}\mathcal{C}^{\mu\nu\rho\eta}_{\se\sz\sd\sv}\left(t,t'\right)
=
\big(\otimes_{\xi} \bra{\Psi_\xi}\big)\hat{a}_{\eta\sv}^{{\rm gr}\,\dagger}(t) \hat{a}_{\eta\sv}^{{\rm ex}}(t) \hat{a}_{\rho\sd}^{{\rm ex}\,\dagger}(t') \hat{a}_{\rho\sd}^{{\rm gr}}(t')\\
\times \hat{a}_{\mu\se}^{{\rm gr}\,\dagger}(t')\hat{a}_{\mu\se}^{{\rm ex}}(t')\hat{a}_{\nu\sz}^{{\rm ex}\,\dagger}(t)\hat{a}_{\nu\sz}^{{\rm gr}}(t)
\big(\otimes_{\chi} \ket{\Psi_\chi}\big)
\,.
\end{multline}
First of all, if there is one index $\eta$, $\rho$, $\mu$ or $\nu$ which does 
not match any of the other three, the expectation value is zero because then
there is always (e.g.)\ a $\hat{a}_{\eta\sv}^{{\rm ex}}$-annihilation operator working on a 
single-site state, which, by assumption, 
contains no  $\hat{a}_{\eta\sv}^{{\rm ex}\,\dagger}$-excitations, i.e.,
\begin{align}
\bra{\Psi_\eta} \hat{a}_{\eta\sv}^{{\rm gr}\,\dagger}(t) \hat{a}_{\eta\sv}^{{\rm ex}}(t) \ket{\Psi_\eta} = 0
\,.
\end{align}
Thus we are left with four possibly nonzero cases -- 
three with two pairs and one where all indices
are the same. 
While the case $\mathcal{C}^{\mu\mu\eta\eta}$ 
was already discussed in Section~\ref{sec:SeparableState},
the cases $\mathcal{C}^{\rho\eta\rho\eta}$ and
$\mathcal{C}^{\eta\rho\rho\eta}$ are zero because, e.g.,
\begin{multline}
\hspace{-0.2cm}\mathcal{C}^{\eta\rho\rho\eta}_{\se\sz\sd\sv}\left(t,t'\right)
=\bra{\Psi_\eta}\bra{\Psi_\rho}\hat{a}_{\eta\sv}^{{\rm gr}\,\dagger}(t) \hat{a}_{\eta\sv}^{{\rm ex}}(t) \hat{a}_{\rho\sd}^{{\rm ex}\,\dagger}(t') \hat{a}_{\rho\sd}^{{\rm gr}}(t')\\
\times\hat{a}_{\eta\se}^{{\rm gr}\,\dagger}(t')\hat{a}_{\eta\se}^{{\rm ex}}(t')\ket{\Psi_\eta}\hat{a}_{\rho\sz}^{{\rm ex}\,\dagger}(t)\hat{a}_{\rho\sz}^{{\rm gr}}(t)\ket{\Psi_\rho}=0
\,.
\end{multline}
When all four indices are the same, $\mathcal{C}^{\eta\eta\eta\eta}$, 
the result is (in general) nonzero. 
It can be neglected, however, as for this case the operatorial
part~(\ref{eq:OpD}) at most scales with $N$ -- as opposed to 
$N^2$ in the case of $\mathcal{C}^{\mu\mu\eta\eta}$.
Hence, to leading order in $N$, the operatorial part
for separable states is given by equation~(\ref{eq:DLowCorr}).
\section{Emission probability density for a single atom}
\label{ap:ProbSingle}
Throughout this paper, we often compare the collective emission
probability density from the atoms in the optical lattice
to the emission probability in case of a single atom.
The special case that there is only a single
atom (without spin quantum number, i.e., index $s$ omitted) 
$\ket{\Psi} = \ket{1}^{{\rm gr}}\otimes\ket{0}^{{\rm ex}}$
absorbing and re-emitting the photon corresponds
to an operatorial part in equation~(\ref{eq:ProbDens2}) of just
\begin{align}
&\hat{\Sigma}^-(\Kout,t_2) \hat{\Sigma}^+(\Kin,t_1)=\nonumber\\
&
\hat{a}^{{\rm gr}\,\dagger}(t_2)\hat{a}^{{\rm ex}}(t_2)\hat{a}^{{\rm ex}\,\dagger}(t_1)\hat{a}^{{\rm gr}}(t_1) \exp\left\{-i\left(\Kout-\Kin\right)\f{r}\right\}=\nonumber\\
&
\hat{a}^{{\rm gr}\,\dagger}\hat{a}^{{\rm ex}}\hat{a}^{{\rm ex}\,\dagger}\hat{a}^{{\rm gr}}\,e^{i\omega (t_1-t_2)} \exp\left\{-i\left(\Kout-\Kin\right)\f{r}\right\}
\,,
\end{align}
because 
$\hat{H}_{\rm lat} = 0$ and thus, $\hat{H}-\hat{V}$ becomes
trivial. Accordingly, $P$ reduces to $P_{\rm single}$ 
(assuming resonance $\omega_{\rm in}=\omega$),
\begin{align}
\label{eq:Psingle}
P_{\rm single}
=
\mathcal{I}\left[ e^{i(\omega_{\rm out}-\omega)t_2} \right]
\,.
\end{align}
Not surprisingly, the emission in case of a single
atom is not directed. 
Note that we assume that the emission
probability of a single atom is independent
of a potential spin quantum number $s$. 
We would expect
the same emission probability density~(\ref{eq:Psingle})
in the case of a single fermionic atom, irregardless
of its spin state.
\section{General lattice Hamiltonian in $\f{k}$-basis}
\label{ap:kBasis}
In the limiting case of weak interactions ($J\gg U$), 
we approximate $U=0$, i.e., the on-site repulsion term vanishes 
from the general lattice Hamiltonian~(\ref{eq:GenHam}),
and only the tunneling term remains,
\begin{align}
\label{eq:HJext}
\hat{H}_{J} =  -\frac{J}{Z}\sum_{\mu\nu, s,\lambda}\tmn \hat{a}_{\mu s}^{\lambda\,\dagger} \hat{a}_{\nu s}^{\lambda}
\,.
\end{align}
The Hamiltonian can be diagonalized using
a Fourier transform to the $\f{k}$-basis,
where the summation runs over all lattice vectors
$\f{k} = \left\{ n, m \right\} \frac{2\pi}{L \ell}$, with 
$n, m \in \left\{-L/2+1,..,L/2\right\}$, 
where $L$ is the number of lattice sites per dimension ($L^2=N$),
\begin{align}
\label{eq:TrafoKspace}
\hat{a}_{\mu s}^{\lambda\,\dagger} &= \frac{1}{\sqrt{N}}\sum_{\fk{k}} \hat{a}_{\fk{k},s}^{\lambda\,\dagger}\exp\left\{-i\f{k}\cdot\f{r}_\mu\right\}
\,.
\end{align}
Applying these on the Hamiltonian~(\ref{eq:HJext}),
we get at first
\begin{multline}
\hat{H}_{J}
=
-\frac{J}{N Z}\sum_{\mu\nu\fk{k}\fk{p},s,\lambda}T_{\mu\nu}
\hat{a}_{\fk{k},s}^{\lambda\,\dagger} \hat{a}_{\fk{p},s}^{\lambda}\\
\times
\exp\left\{-i\f{k}\cdot\f{r}_\mu\right\}
\exp\left\{i\f{p}\cdot\f{r}_\nu\right\}
\,.
\end{multline}
However, for a 2D-lattice, the summation over 
the four nearest neighbors yields
\begin{align}
\label{eq:SummNN}
\sum_{\nu}T_{\mu\nu} \exp\left\{i\f{p}\cdot\f{r}_\nu\right\}
= T_{\fk{p}}\exp\left\{i\f{p}\cdot\f{r}_\mu\right\}
\,,
\end{align}
with $T_{\fk{p}}$ the Fourier transform of the 
adjacency matrix,
\begin{align}
T_{\fk{p}} = 2\left[\cos(p_x \ell) + \cos(p_y \ell)\right]
\,.
\end{align}
In addition, as $\f{k}$ and $\f{p}$ are both lattice vectors, the
$\mu$-summation is non-vanishing only for $\f{k}=\f{p}$,
i.e.,
\begin{align}
\label{eq:SummDelta}
\sum_{\mu} \exp\left\{i\left(\f{p}-\f{k}\right)\cdot\f{r}_\mu\right\} = N \delta_{\fk{k}\fk{p}}
\,.
\end{align}
Employing the relations~(\ref{eq:SummNN}) and%
~(\ref{eq:SummDelta}), we arrive at the diagonalized 
tunneling Hamiltonian in $\f{k}$-basis
\begin{align}
\label{eq:HJextk}
\hat{H}_{J}
=
-\frac{J}{Z}\sum_{\fk{k},s,\lambda}T_{\fk{k}}^{\phantom{\,}}
\hat{a}_{\fk{k},s}^{\lambda\,\dagger}\hat{a}_{\fk{k},s}^{\lambda}
=
-\frac{J}{Z}\sum_{\fk{k},s,\lambda}T_{\fk{k}}^{\phantom{\,}}
\hat{n}_{\fk{k},s}^{\lambda}
\,.
\end{align}
\section{Four-point correlators in the bosonic and fermionic case}
\label{ap:FourPointCorr}
\subsection{Correlator in the bosonic case}
The most explanatory way to calculate the bosonic four-point correlator 
in equation~(\ref{eq:OpE}),
\begin{align}
E_n^{\rm B} = \bra{\Psi}\hat{a}_{\fk{q}-\kin}^{{\rm gr}\,\dagger}\hat{a}_{\fk{q}-\kout}^{{\rm gr}} \hat{a}_{\fk{k}-\kout}^{{\rm gr}\,\dagger}\hat{a}_{\fk{k}-\kin}^{{\rm gr}}\ket{\Psi}
\,,
\end{align}
is by going through the cases which can yield a nonzero result. 
Obviously, the two creation operators have to compensate for the two
annihilation operators, such that in the end, the basis-state $\ket{\Psi}$
is unchanged.
This is the case when either  
$\f{k}-\Kin=\f{k}-\Kout$ and $\f{q}-\Kout=\f{q}-\Kin$,
or $\f{k}-\Kin=\f{q}-\Kin$ and $\f{k}-\Kout=\f{q}-\Kout$, i.e.,
\begin{align}
E_n^{\rm B} &=\delta_{\fk{k}-\kin,\fk{k}-\kout} \delta_{\fk{q}-\kout,\fk{q}-\kin} \bra{\Psi}\hat{n}_{\fk{q}-\kout}^{{\rm gr}}\hat{n}_{\fk{k}-\kin}^{{\rm gr}}\ket{\Psi}\nonumber\\
&\hspace{0.4cm}+
\delta_{\fk{k}-\kin,\fk{q}-\kin} \delta_{\fk{k}-\kout,\fk{q}-\kout}\nonumber\\
&\hspace{0.9cm}\times\bra{\Psi} \hat{a}_{\fk{k}-\kin}^{{\rm gr}\,\dagger}\hat{a}_{\fk{q}-\kout}^{{\rm gr}}\hat{a}_{\fk{q}-\kout}^{{\rm gr}\,\dagger}\hat{a}_{\fk{k}-\kin}^{{\rm gr}}\ket{\Psi}\nonumber\\
&\hspace{0.9cm}\times\left(1-\delta_{\fk{k}-\kin,\fk{q}-\kout}\right)
\,,
\end{align}
where the case that all four operators work on the same $\f{k}$-vector
was excluded in the last line, as it is already included in the first term.
Further simplifications yield
\begin{align}
\label{eq:Corr_Bos_Res}
E_n^{B} 
&=\delta_{\kin\kout}\bra{\Psi}\hat{n}_{\fk{q}-\kout}^{{\rm gr}}\hat{n}_{\fk{k}-\kin}^{{\rm gr}}\ket{\Psi}\\
&\hspace{0.4cm}+\delta_{\fk{k}\fk{q}} \bra{\Psi}\left(\hat{n}_{\fk{q}-\kout}^{{\rm gr}}+1\right)\hat{n}_{\fk{k}-\kin}^{{\rm gr}}\ket{\Psi}\left(1-\delta_{\kin\kout}\right)\nonumber
\,.
\end{align}
And with $\hat{n}_{\fk{k}}^{{\rm gr}}\ket{\Psi} = n(\f{k}) \ket{\Psi}$,
where $\ket{\Psi}$ is of course normalized,
we arrive at the result
\begin{align}
\label{eq:Corr_Bos_Res_Num}
E_n^{\rm B} 
&=n(\f{k}-\Kin) n(\f{q}-\Kout)\left(\delta_{\kin\kout}+\delta_{\fk{k}\fk{q}}\right)\nonumber\\
&\hspace{0.4cm}+n(\f{k}-\Kin)\delta_{\fk{k}\fk{q}}\nonumber\\
&\hspace{0.4cm}-n(\f{k}-\Kin) \left[ n(\f{q}-\Kout) + 1 \right] \delta_{\fk{k}\fk{q}}\delta_{\kin\kout}
\,.
\end{align}
\subsection{Correlator in the fermionic case}
The fermionic correlator from equation~(\ref{eq:OpE}),
\begin{align}
\label{eq:Corr_Ferm}
E_n^{\rm F} = \bra{\Psi}\hat{a}_{\fk{q}-\kin,\sz}^{{\rm gr}\,\dagger}\hat{a}_{\fk{q}-\kout,\sz}^{{\rm gr}} \hat{a}_{\fk{k}-\kout,\se}^{{\rm gr}\,\dagger}\hat{a}_{\fk{k}-\kin,\se}^{{\rm gr}}
\ket{\Psi}
\,,
\end{align}
can be calculated analogous to the bosonic case,
although some differences arise.
Again, a nonzero result is obtained when 
$\f{k}-\Kin = \f{k}-\Kout$, $\f{q}-\Kout = \f{q}-\Kin$
and spins $\se,\sz$ arbitrary.
The other possible combination is when
$\f{k}-\Kin=\f{q}-\Kin$, $\f{k}-\Kout=\f{q}-\Kout$
and the spins $\se=\sz$ are equal,
\begin{align}
\label{eq:Corr_Cases_Ferm}
E_n^{\rm F} &=\delta_{\fk{k}-\kin,\fk{k}-\kout} \delta_{\fk{q}-\kout,\fk{q}-\kin}\nonumber\\
&\hspace{0.4cm}\times \bra{\Psi} \hat{n}_{\fk{q}-\kout,\sz}^{{\rm gr}} \hat{n}_{\fk{k}-\kin,\se}^{{\rm gr}} \ket{\Psi}\nonumber\\
&\hspace{0.2cm}+
\delta_{\fk{k}-\kin,\fk{q}-\kin} \delta_{\fk{k}-\kout,\fk{q}-\kout}\delta_{\se\sz}\left(1-\delta_{\fk{k}-\kin,\fk{q}-\kout}\right)\nonumber\\
&\hspace{0.4cm}\times \bra{\Psi}\hat{a}_{\fk{k}-\kin,\se}^{{\rm gr}\,\dagger}\hat{a}_{\fk{q}-\kout,\se}^{{\rm gr}}\hat{a}_{\fk{q}-\kout,\se}^{{\rm gr}\,\dagger}\hat{a}_{\fk{k}-\kin,\se}^{{\rm gr}} \ket{\Psi}
\,,
\end{align}
where, again, the case of all four $\f{k}$-vectors equal
was excluded from the second term to prevent it being
included twice.
After simplifying the Kronecker deltas and applying anti-commutation
relations to the second term, we arrive at:
\begin{align}
E_n^{\rm F} &=\delta_{\kin \kout}\bra{\Psi} \hat{n}_{\fk{q}-\kout,\sz}^{{\rm gr}} \hat{n}_{\fk{k}-\kin,\se}^{{\rm gr}} \ket{\Psi}\nonumber\\
&\hspace{0.4cm}+
\delta_{\fk{k} \fk{q}}\delta_{\se\sz}\left(1-\delta_{\kin \kout}\right)\nonumber\\
&\hspace{0.9cm}\times \bra{\Psi}\left(1-\hat{n}_{\fk{q}-\kout,\se}^{{\rm gr}}\right)\hat{n}_{\fk{k}-\kin,\se}^{{\rm gr}} \ket{\Psi}
\,.
\end{align}
This expression resembles the result of the bosonic case 
(\ref{eq:Corr_Bos_Res}), but is not quite the same.
The difference becomes even more apparent when
we consider that we can omit the $\delta_{\kin \kout}$ in the second line,
as in the fermionic case $(\hat{n}_{\fk{k}-\kin,\se}^{{\rm gr}})^2 = \hat{n}_{\fk{k}-\kin,\se}^{{\rm gr}}$,
i.e., the expectation value already becomes zero for $\Kin=\Kout$ (and $\f{k}=\f{q}$),
\begin{align}
E_n^{\rm F} &=\delta_{\kin \kout}\bra{\Psi} \hat{n}_{\fk{q}-\kout,\sz}^{{\rm gr}} \hat{n}_{\fk{k}-\kin,\se}^{{\rm gr}} \ket{\Psi}\nonumber\\
&\hspace{0.4cm}+\delta_{\fk{k} \fk{q}}\delta_{\se\sz} \bra{\Psi}\left(1-\hat{n}_{\fk{q}-\kout,\se}^{{\rm gr}}\right)\hat{n}_{\fk{k}-\kin,\se}^{{\rm gr}} \ket{\Psi}
\,.
\end{align}
Expressed via the occupation numbers $\hat{n}_{\fk{k},s}^{{\rm gr}}\ket{\Psi} = n_s(\f{k}) \ket{\Psi}$, we come to the final result
\begin{align}
\label{eq:Corr_Ferm_Res_Num}
E_n^{\rm F} &=
n_{\se}\left(\f{k}-\Kin\right) n_{\sz}\left(\f{q}-\Kout\right) \left( \delta_{\kin \kout} - \delta_{\fk{k} \fk{q}}\delta_{\se\sz} \right)\nonumber\\
&\hspace{0.4cm}+n_{\se}\left(\f{k}-\Kin\right) \delta_{\fk{k} \fk{q}} \delta_{\se\sz}
\,.
\end{align}
\section{Emission probability of the partial condensation state}
\label{ap:PartialCond}
In this section, we derive the emission probability density of the
partial condensation state~(\ref{eq:probability}), which contains 
a certain number $N_1$ of bosons in the ground state at $\f{k}=0$ 
while the other $N_2$ bosons should be equally distributed 
between all $\f{k}$-modes%
\footnote{
Strictly speaking, the presented partial condensation
state is neither a pure state diagonal in the $\f{k}$-basis
nor a Gaussian state, when $N_{1/2}$ are chosen such
that $N_2/N$ is not an integer.
However, there are two possible ways out, such that the 
partial condensation state can be expressed as a pure state.
One could either consider a lattice with more atoms $N_\Psi$ 
than lattice sites $N$ to make sure that $N_2/N$ is integer, or
one can think of distributing the $N_2$ bosons only 
almost equally (e.g., checkerboard pattern), 
such that the number of bosons in
a certain $\f{k}$-mode is always integer.
}.
Inserting the corresponding number distribution $n^{\rm dt}(\f{k}) =
N_1\delta_{\fk{k}\fk{0}} + N_2/N$ into the expression for the 
bosonic four-point correlator~(\ref{eq:BosonicCorr}) yields a 
lengthy result,
\begin{align}
\label{eq:CorrNdt}
E_{n^{\rm dt}}^{\rm B} &=
N_1^2\delta_{\fk{k}\fk{q}\kin\kout}\nonumber\\
&\hspace{0.4cm}+\frac{N_1 N_2}{N} (\delta_{\fk{k}\kin\kout}+\delta_{\fk{q}\kin\kout}\nonumber\\
&\hspace{1.5cm}+\delta_{\fk{k}\fk{q}\kin}+\delta_{\fk{k}\fk{q}\kout}-2\delta_{\fk{k}\fk{q}\kin\kout})\nonumber\\
&\hspace{0.4cm}+N_1\left(\delta_{\fk{k}\fk{q}\kin}-\delta_{\fk{k}\fk{q}\kin\kout}\right)\nonumber\\
&\hspace{0.4cm}+\frac{N_2^2}{N^2} \left(\delta_{\kin\kout}+\delta_{\fk{k}\fk{q}}-\delta_{\fk{k}\fk{q}}\delta_{\kin\kout}\right)\nonumber\\
&\hspace{0.4cm}+\frac{N_2}{N}\left(\delta_{\fk{k}\fk{q}}-\delta_{\fk{k}\fk{q}}\delta_{\kin\kout}\right)
\,.
\end{align}
However, assuming that $N_1$, $N_2$ and $N$ are of the same 
order $\ord(N)\gg1$, we can omit all terms of order 
$\ord(N)$ or lower from $E_{n^{\rm dt}}^{\rm B}$, 
i.e., keep only terms of quadratic order $\ord(N^2)$. 
Note that the summation over the indices $\f{k}$ and $\f{q}$
in equation~(\ref{eq:OpE}) in general also yields a factor 
of order $\ord(N)$ each, and thus it is important
whether these indices are fixed or free in the respective term.
For example, the term $N_1 N_2/N \delta_{\fk{k}\fk{q}\kin}$ 
in the third line in equation~(\ref{eq:CorrNdt}) only scales with 
$\ord(N)$, due to the prefactor, as both indices 
$\f{k}$ and $\f{q}$ are fixed. This is in opposition to, e.g., the
terms of the second line in equation~(\ref{eq:CorrNdt}).
After omitting all terms of order $\ord(N)$ or lower,
which correspond to incoherent emission processes,
the remaining terms,
\begin{multline}
\label{eq:EndtApprox}
\tilde{E}_{n^{\rm dt}}^{\rm B} 
=
\bigg(N_1^2\delta_{\fk{k}\fk{q}\kout}
+ \frac{N_1 N_2}{N} \left(\delta_{\fk{k}\kout}+\delta_{\fk{q}\kout}\right)+
\frac{N_2^2}{N^2} \bigg)\\
\times\, \delta_{\kin\kout}
\,,
\end{multline}
carry the factor $\delta_{\kin\kout}$,
that is, they represent superradiant emission in the same 
direction in which the absorbed photon was incoming,
$\Kout=\Kin$.
When we insert~(\ref{eq:EndtApprox})
into the operatorial part $\mathcal{E}$ in equation~(\ref{eq:OpE}),
we can thus combine the phases in the 
exponentials. After we have introduced 
$\Ka = \Kin = \Kout$, 
$\varphi(\Delta t) = -\phi_{\ka}^{\ka}(\Delta t)$
and the phase-sum $\mathcal{J}$ as abbreviations,
\begin{align}
\label{eq:reduction}
\mathcal{J}(\Delta t)
=
\frac1N\sum_{\fk{p}}
\exp\left\{iJ/Z(T_{\fk{p}}-T_{\fk{p}-\fk{\kappa}})\Delta t\right\}
\leq1
\,,
\end{align}
the operatorial part reads
\begin{align}
\mathcal{E}\left(t_1, t_2, t_3, t_4\right)
&=
\Bigg[
N_1^2 
e^{i \varphi(t_3-t_4)}
e^{i \varphi(t_2-t_1)}\nonumber\\
&\hspace{-2.0cm}+N_1 N_2
\left(
\mathcal{J}(t_3-t_4)
e^{i\varphi(t_2-t_1)}
+\mathcal{J}(t_2-t_1)
e^{i\varphi(t_3-t_4)}
\right)\nonumber\\
&\hspace{-0.1cm}+N_2^2 \mathcal{J}(t_3-t_4)\mathcal{J}(t_2-t_1)
\Bigg] \delta_{\kin\kout}
\,.
\end{align}
Finally, we can evaluate the emission probability
density (\ref{eq:ProbDensHigh}). It can be written
as an absolute square depending on the waiting
time $\Delta t=t_2-t_1$ times the single-atom 
emission probability density $P_{\rm single}$
(when we again regard $\Delta t$ as constant
over the integration periods, see above),
\begin{align}
P=\left|N_1 e^{i\varphi(\Delta t)}+N_2{\mathcal J}(\Delta t)\right|^2
\delta_{\kin\kout}P_{\rm single}
\,.
\end{align}
\section{Approximation of the phase-sum $\mathcal{J}(\Delta t)$ via Bessel functions}
\label{ap:approx_phasesum}
The phase-sum~(\ref{eq:reduction}),
\begin{align}
\label{eq:ap:reduction}
\mathcal{J}(\Delta t)
=
\frac1N\sum_{\fk{p}}
\exp\left\{iJ/Z(T_{\fk{p}}-T_{\fk{p}-\fk{\kappa}})\Delta t\right\}
\,,
\end{align}
can be approximated when the photon wave number $\Ka$ is small
compared to the inverse of the lattice constant $\ell$.
In this case, we can taylor-approximate the argument of the
exponential,
\begin{align}
\label{eq:ap:ApproxSmallKappa}
\left(T_{\fk{p}}-T_{\fk{p}-\ka}\right) &\approx \Ka\cdot\nabla_{\fk{p}}T_{\fk{p}} \big|_{\f{p}}\nonumber\\
&= -2 \ell\left[ \kappa_x \sin\left(p_x \ell\right) + \kappa_y \sin\left(p_y \ell\right)\right]
\,.
\end{align}
And split the phase-sum $\mathcal{J}$ in equation~(\ref{eq:ap:reduction}) into
a $x$- and a $y$-part,
\begin{align}
\mathcal{J}(\Delta t)
=
\mathcal{J}^{x}(\Delta t)\times\mathcal{J}^{y}(\Delta t)
\,,
\end{align}
with the one-dimensional sums over the $L$ values of $p_{x}$ or $p_{y}$
(where $L$ is the number of lattice sites in one dimension, i.e., $L^2=N$),
\begin{align}
\label{eq:ap:Jxy}
\mathcal{J}^{x/y}(\Delta t):=\frac{1}{L}\sum_{p_{x/y}} \exp\left\{-i2\ell J/Z \kappa_{x/y} \sin(p_{x/y} \ell)\Delta t\right\}
\,.
\end{align}
These one-dimensional sums can now be approximated as (cylindrical) Bessel functions when going to the integral (justified for $L\gg1$),
\begin{multline}
\mathcal{J}^{x/y}(\Delta t)
=
\frac{1}{L}\hspace{-0.9cm}\sum_{\hspace{1.0cm}n=-L/2+1}^{L/2}\hspace{-0.9cm}\exp\left\{-i2\ell J/Z \kappa_{x/y} \sin\left( n 2\pi / L\right)\Delta t\right\}\\
\approx \frac{1}{L}\int_{-L/2}^{L/2} d\lambda\,\exp\left\{-i2\ell J/Z \kappa_{x/y} \sin\left( \lambda 2\pi / L\right)\Delta t\right\} 
\,,
\end{multline}
and substituting by $z = \lambda 2 \pi / L$,
\begin{align}   
\mathcal{J}^{x/y}(\Delta t)
&\approx
\frac{1}{2\pi}\int_{-\pi}^{\pi} dz\,\exp\left\{-i2\ell J/Z \kappa_{x/y} \sin\left(z\right)\Delta t\right\}\nonumber\\
&= J_0 \left( -2\frac{J \Delta t}{Z}\kappa_{x/y} \ell \right)
\,.
\end{align}
Thus, in summary, the phase-sum~(\ref{eq:reduction}) can be approximated via
\begin{align}
\mathcal{J}(\Delta t)
\approx
J_0 \left( 2\frac{J\Delta t}{Z}\,\kappa_x\ell \right) 
J_0 \left( 2\frac{J\Delta t}{Z}\,\kappa_y\ell \right) 
\,,
\end{align}
where the minus signs were dropped as the Bessel functions
$J_0$ are even.
\section{Four-point correlators in the Mott state and Mott-N\'{e}el state}
\label{ap:FourPointCorrMott}
To calculate the four-point correlator in the 
Mott state~(\ref{eq:MottBos}) or the
Mott-N\'{e}el state~(\ref{eq:NeelFerm}),
\begin{multline}
E_{\mon}^{\rm B/F}
=
\prescript{J=0}{\mon}{\bra{\Psi}}
\hat{a}_{\fk{q}-\kin,\sz}^{{\rm gr}\,\dagger}\hat{a}_{\fk{q}-\kout,\sz}^{{\rm gr}}\\
\times\hat{a}_{\fk{k}-\kout,\se}^{{\rm gr}\,\dagger}\hat{a}_{\fk{k}-\kin,\se}^{{\rm gr}}
\ket{\Psi}_{\mon}^{J=0}
\,,
\end{multline}
we transfer the creation and annihilation
operators back to the lattice site basis via
\begin{align}
\label{eq:TrafoLattice}
\hat{a}_{\fk{k},s}^{\lambda\,\dagger} 
= 
\frac{1}{\sqrt{N}}\sum_{\mu}\hat{a}_{\mu s}^{\lambda\,\dagger}\exp\left\{i\f{k}\cdot\f{r}_\mu\right\}
\,,
\end{align}
leading to the expression
\begin{align}
\label{eq:CorrMottLattice}
E_{\mon}^{\rm B/F}
&=
\frac{1}{N^2}\sum_{\mu\nu\rho\eta}
e^{i(\fk{k}-\kout)\fk{r}_\mu} e^{-i(\fk{k}-\kin)\fk{r}_\nu}\nonumber\\
&\hspace{1.3cm}\times e^{-i(\fk{q}-\kout)\fk{r}_\rho} e^{i(\fk{q}-\kin)\fk{r}_\eta}\nonumber\\
&\hspace{-1.0cm}\times
\prescript{J=0}{\mon}{\bra{\Psi}}\hat{a}_{\eta\sz}^{{\rm gr}\,\dagger} \hat{a}_{\rho\sz}^{{\rm gr}} \hat{a}_{\mu\se}^{{\rm gr}\,\dagger}\hat{a}_{\nu\se}^{{\rm gr}} \ket{\Psi}_{\mon}^{J=0}
\,.
\end{align}
As now both the operators and the states are given in the lattice
site basis, the expectation value in the last line can be readily
calculated.
For the case of bosons (i.e., index $s$ omitted) and
the Mott insulator state, we get
\begin{multline}
\prescript{J=0}{\mot}{\bra{\Psi}}\hat{b}_{\eta}^{{\rm gr}\,\dagger}\hat{b}_{\rho}^{{\rm gr}}\hat{b}_{\mu}^{{\rm gr}\,\dagger}\hat{b}_{\nu}^{{\rm gr}}\ket{\Psi}_{\mot}^{J=0}
=\\
\delta_{\mu\nu}\delta_{\rho\eta} + 2\delta_{\nu\eta}\delta_{\mu\rho} - 2\delta_{\mu\nu\rho\eta}
\,,
\end{multline}
and thus
\begin{align}
E_{\mot}^{\rm B} 
=
\delta_{\kin \kout} + 2 \delta_{\fk{k} \fk{q}} - 2/N
\,.
\end{align}
For the case of fermions and the Mott-N\'{e}el state,
we get
\begin{multline}
\prescript{J=0}{\nee}{\bra{\Psi}}\hat{c}_{\eta\sz}^{{\rm gr}\,\dagger}\hat{c}_{\rho\sz}^{{\rm gr}} \hat{c}_{\mu\se}^{{\rm gr},\dagger}\hat{c}_{\nu\se}^{{\rm gr}}\ket{\Psi}_{\nee}^{J=0}
=
\delta_{\mu\nu}\delta_{\rho\eta}\delta_{\se s_\nu}\delta_{\sz s_\rho}\\
 + \delta_{\nu\eta}\delta_{\mu\rho}\delta_{\se\sz s_\nu} - \delta_{\nu\eta}\delta_{\mu\rho} \delta_{\se\sz s_\nu s_\mu}
\,,
\end{multline}
where the $s_{\nu}$ refer to the spin of the atom 
at lattice site $\nu$ in the Mott-N\'{e}el state, and so forth.
Inserting in~(\ref{eq:CorrMottLattice}), and summing over
the spin indices $\se, \sz$, yields
\begin{multline}
\sum_{\se\sz} E_{\nee}^{\rm F}
=
\delta_{\kin\kout}+\delta_{\fk{k}\fk{q}}\\
-
\frac{1}{N^2}\sum_{\mu\nu}
e^{i(\fk{k}-\fk{q})\fk{r}_\mu} e^{-i(\fk{k}-\fk{q})\fk{r}_\nu}
\delta_{s_\nu s_\mu}
\,.
\end{multline}
The third term can be calculated by splitting the sums
according to the two sub-lattices $\mathcal{A}$ and $\mathcal{B}$
(see above),
which yields two contributions with $(N/2)^2$ summands each, i.e.,
together they give $N^2\delta_{\fk{k}\fk{q}}/2$, and thus,
\begin{align}
\sum_{\se\sz} E_{\nee}^{\rm F} 
=
\delta_{\kin\kout} + \delta_{\fk{k}\fk{q}}/2 
\,.
\end{align}
\section{Adiabatic transition in a bosonic lattice}
\label{ap:AdiabTransBos}
In the bosonic case, the excited state%
~(\ref{eq:MottNeelEx}) reads
\begin{align}
\label{eq:Mott-excited}
\ket{\Psi}_{\exc}^{\mot}
\propto
\frac{1}{\sqrt{N}}\sum_{\mu} \exp\left\{i\Kin\cdot\f{r}_\mu\right\}\hat{b}_{\mu}^{{\rm ex}\,\dagger}\prod\limits_{\nu\neq\mu}\hat{b}_{\nu}^{{\rm gr}\,\dagger}\ket{0}
\,,
\end{align}
where the $\ket{0}$-kets refers to the vacuum state of both the 
lattice bosons and the photon field.
Note that after the absorption of the probe photon $\Kin$,
the lattice state lives in the subspace where 
one atom is excited and the other $N-1$ atoms are not,
and stays in that subspace during the adiabatic evolution
and the following waiting time, i.e., until the probe photon
is re-emitted.
Moreover, in the separable state regime ($U\gg J$),
the state~(\ref{eq:Mott-excited}) is an approximate eigenstate of the
two-species Bose-Hubbard Hamiltonian~(\ref{eq:BHHamExt}), 
when $\Kin$ is a lattice vector (what we require).
We know from the adiabatic theorem that an initial eigenstate
such as~(\ref{eq:Mott-excited}) stays an eigenstate during the
adiabatic evolution, and we end up with a corresponding
eigenstate of the new ($U=0$) Hamiltonian.
For $N\gg 1$, this corresponding eigenstate 
(with the same prefactor stemming from the absorption) 
is given by
\begin{align}
\label{eq:sf-excited}
\ket{\Psi}_{\exc}^{\sfl}
\propto
\frac{1}{\sqrt{(N-1)!}}\,
\big(\hat{b}_{\fk{k}=0}^{{\rm gr}\,\dagger}\big)^{N-1}\,\hat{b}_{\kin}^{{\rm ex}\,\dagger}
\ket{0}
\,,
\end{align}
i.e., with $N-1$ atoms in the superfluid state at $\f{k}=0$,
while the excited atom carries the wave vector $\Kin$ of the
absorbed photon.
The emission probability after the adiabatic transition
can be calculated analog to~(\ref{eq:PMottexcited}).
As a shortcut, the state~(\ref{eq:sf-excited}) can be understood
as exciton creation applied to the superfluid ground state, i.e.,
the emission probability can be evaluated via
\begin{multline}
\hat{\Sigma}_J^-\left(\Kout,t_2\right)
\hat{\Sigma}_U^+(\Kin,t_1)\ket{\Psi}_{\sfl}^{U=0}
=e^{i\omega (t_1-t_2)}\\
\times
\sum_{\fk{k}} \exp\left\{-i\phi_{\fk{k}}^{\kout}(t_2)\right\}
\hat{b}_{\fk{k}-\kout}^{{\rm gr}\,\dagger}\hat{b}_{\fk{k}-\kin}^{{\rm gr}}\ket{\Psi}_{\sfl}^{U=0}
\,,
\end{multline}
which leads back to the already known four-point correlator
of the superfluid ground state~(\ref{eq:CorrSfGround}).
Connecting the parts we obtain in leading order in $N$
\begin{align}
P
=
N^2 \delta_{\kin\kout}\,\mathcal{I}\left[ e^{i(\omega_{\rm out}-\omega)t_2}e^{i\varphi(t_2)} \right]
\,,
\end{align}
where, thanks to the properties of the superfluid ground state,
the $\f{k}$-sums are fixed and only a single phase factor
$e^{i\varphi(t_2)}$ remains.
This phase can be neglected, however, as
the tunneling rate $J$ is small compared to the optical 
energies in the temporal phase.
Thus we arrive at the result of usual, full superradiance,
\begin{align}
\label{eq:PadiabBos}
P = N^2 \delta_{\kin\kout} P_{\rm single}
\,.
\end{align}


\bibliographystyle{epj}
\bibliography{l9}

\end{document}